\documentclass[11pt, a4paper]{article}
\usepackage{styleBuding}
\usepackage{bm}
\usepackage{color}
\usepackage[usenames,dvipsnames,svgnames,table]{xcolor}
\usepackage{amsmath}
\usepackage{amssymb}
\usepackage{graphicx}
\usepackage{slashed}
\usepackage{soul}
\usepackage{multirow}
\usepackage{subfigure}
\usepackage{booktabs}
\usepackage{siunitx} 

\DeclareMathOperator{\sgn}{sgn}
\pdfoutput=1

\title{\boldmath Anomalous Muon Magnetic Moment in the Inverse Seesaw Extended Next-to-Minimal Supersymmetric Standard Model}

\author{Junjie Cao,}
\author{Jingwei Lian,}
\author{Lei Meng,}
\author{Yuanfang Yue,}
\author{and Pengxuan Zhu}

\affiliation{Department of Physics, Henan Normal University, Xinxiang 453007, China}

\emailAdd{junjiec@alumni.itp.ac.cn}
\emailAdd{ljwfly@hotmail.com}
\emailAdd{mel18@foxmail.com}
\emailAdd{yuanfang405@gmail.com}
\emailAdd{zhupx99@icloud.com}

\abstract{The present work investigates the possibility that both dark matter and the anomalous magnetic moment of the muon may be explained within the context of the inverse seesaw extended Next-to-Minimal Supersymmetric Standard Model (ISS-NMSSM). In ISS-NMSSM, the newly introduced Higgs-neutrino Yukawa coupling $Y_\nu$ provides additional Higgsino-sneutrino loop contribution to $(g-2)_{\mu}$. If the deviation between the experimental observations and the Standard Model predictions of the anomalous muon magnetic moment is confirmed by the further experimental and theoretical studies, it can be explained naturally within the ISS-NMSSM framework without conflicting with the current stringent limits on the direct detection of dark matter and Large Hadron Collider searches.}

\begin{document}
\maketitle
\section{Introduction}
\label{sec:intro}
The anomalous magnetic moment of the muon $a_{\mu}$ represents a rigorous test of the Standard Model (SM) of particle physics. Here, a $3.5~\sigma$ discrepancy has been observed between the high precision experimental measurements of $a_{\mu}$ achieved by the Brookhaven National Laboratory (BNL) E821 experiment~\cite{Bennett:2006fi} and its theoretical calculation based on the SM~\cite{Blum:2013xva, Keshavarzi:2018mgv, Aoyama:2012wk, Aoyama:2017uqe, Gnendiger:2013pva, Nyffeler:2016gnb, Davier:2019can}. Quantitatively, this discrepancy between the experimentally measured value $a_{\mu}^{\rm exp}$ and the theoretical value $a_{\mu}^{\rm SM}$ is given as \cite{Davier:2019can, Davier:2017zfy, Tanabashi:2018oca}
\begin{equation}\label{eq:damu}
    \Delta a_{\mu} = a_{\mu}^{\rm exp} - a_{\mu}^{\rm SM} = 268(63)(43)\times 10^{-11},
\end{equation} 
where the number in the first parentheses is the current experimental uncertainty from BNL E821, and the second is the total theoretical uncertainty\footnote{The dominating limitation of the theoretical precision comes from the uncertainty of the hadronic vacuum polarization (HVP). The contribution of non-perturbative energy region HVP is evaluated either from low-energy $e^+e^-\to {\rm hadron}$ cross section or from hadronic $\tau$-decay due to an isospin symmetry. In the Eq.~(\ref{eq:damu}), only the $e^+e^-$ annihilation data is included in the estimation of HVP, while the $\tau^-\to \pi^- \pi^0 \nu_\tau$ data is not taken into consideration. Currently, although the error of $e^+e^-$ data is relatively small in comparison with the $\tau$ decay data, the $\tau$ lepton plays an interesting and special role in testing SM and in evaluating the HVP. Specially, the discrepancy $\Delta a_\mu$ based on the $\tau$-data is only $195 \pm 83 \times 10^{-11}$, which deviates from the experimental observation by $2.4~\sigma$ and about $2.2~\sigma$ from the $e^+e^-$ data based prediction~\cite{Davier:2010nc, Davier:2013sfa, Hagiwara:2011af}. Compared with the $e^+e^-$ based result, the $\tau$-decay based result is more supporting the hypothesis that there is no $(g-2)_{\mu}$ deviation \cite{Jegerlehner:2009ry, Beskidt:2012sk, Benayoun:2011mm, Benayoun:2012wc, Benayoun:2015gxa}. }. 
Meanwhile, the upcoming Fermilab E989 experiment~\cite{Grange:2015fou} is expected to improve on the precision of $a_{\mu}^{\rm exp}$ by a factor of four compared to the BNL E821 experiment, and therefore has the capability of providing a more precise test on SM\footnote{A $5\sigma$ discovery level conclusion of $\Delta a_\mu$ not only needs experimental effort, but also the advance in theoretical evaluation accuracy, which requires a two times precision improvement of $a_{\mu}^{\rm SM}$ compared to the current estimation. Moreover, the explanation of the discrepancy between the $\tau$-decay based and the $e^+e^-$ annihilation based extractions of the HVP contribution to $a_{\mu}^{\rm SM}$ also needs further verification. }.

\par The discrepancy $\Delta a_\mu$ also provides an excellent perspective for investigating the physics beyond the SM (BSM). Various BSM mechanisms have been proposed to account for $\Delta a_{\mu}$. To our best knowledge, the proposed mechanisms can be given as follows.
\begin{itemize}
	\item Extra $U(1)$ gauge extension of SM frameworks. These include the dark $Z$ model~\cite{Altmannshofer:2016brv} and the lepton flavor violating $U(1)_{L_\mu - L_\tau}$ model~\cite{Krnjaic:2019rsv, Darme:2018hqg, Altmannshofer:2014pba, Foldenauer:2018zrz} with an ${L_\mu - L_\tau}$ gauge symmetry and $m_{Z^{\prime}} >  m_{\tau} - m_{\mu}$, where $m_{Z^{\prime}}$ is the mass of the ${Z^\prime}$ gauge boson, $m_{\tau}$ is the $\tau$ lepton mass, and $m_{\mu}$ is the muon mass. These frameworks can account for $\Delta a_{\mu}$ under current experimental constraints. In contrast, the dark photon model~\cite{Davoudiasl:2014kua, Mohlabeng:2019vrz, Fuyuto:2019vfe}, $U(1)_{B-L}$ model, and $U(1)_{L_\mu-L_e}$ model fail to explain $\Delta a_{\mu}$ under current experimental constraints~\cite{Bauer:2018onh}.
	\item Two Higgs doublet model (2HDM) frameworks. These include the aligned 2HDM with light neutral Higgs bosons of masses $3~{\rm GeV} \leq m_{H} \leq 50~{\rm GeV}$ and $10~{\rm GeV} \leq m_{A} \leq 130~{\rm GeV}$~\cite{Han:2015yys, Ilisie:2015tra, Cherchiglia:2016eui}, a muon specific 2HDM~\cite{Abe:2017jqo} with an enhancement factor $\tan{\beta} \sim \mathcal{O}(1000)$, where $\tan{\beta}$ is the ratio of the vacuum expectation values of the two doublet Higgs fields, a lepton specific (Type-X) 2HDM~\cite{Cao:2009as, Broggio:2014mna, Wang:2014sda, Abe:2015oca}, a U(1)-symmetric 2HDM~\cite{Li:2018aov}, and a $\mu-\tau$ lepton flavor violating 2HDM~\cite{Iguro:2019sly}.
	\item Supersymmetry frameworks. These include Minimal Supersymmetric SM (MSSM) \cite{Stockinger:2006zn, Martin:2001st, Belyaev:2016oxy}, vector-like extended MSSM \cite{Choudhury:2017fuu,Choudhury:2017acn}, and the Minimal R-symmetric Supersymmetric SM (MRSSM) \cite{Kotlarski:2019muo}\footnote{In contrast to the MSSM, the enhancement factor $\tan{\beta}$ is absent in the MRSSM.}.
\end{itemize}
 Among these frameworks, supersymmetry (SUSY) explanations are quite naturally applied to account for $\Delta a_{\mu}$ because they introduce an additional supersymmetric contribution $a_{\mu}^{\rm SUSY}$ to $a_{\mu}$, which is generically given as follows~\cite{Czarnecki:2001pv}.
\begin{equation}
\begin{split}
    a_{\mu}^{\rm SUSY} &\simeq \sgn(\mu)\frac{\alpha(m_Z)}{8\pi\sin^2{\theta_W}}\frac{m_{\mu}^2}{M_{\rm SUSY}^2}\tan{\beta}\left(  1-\frac{4\alpha}{\pi}\ln{\frac{M_{\rm SUSY}}{m_{\mu}}} \right) \\
    &\simeq \sgn(\mu)~130\times10^{-11}\cdot \left( \frac{100~{\rm GeV}}{M_{\rm SUSY}} \right)^2 \tan{\beta}
    \end{split}
\end{equation}
Here, $\mu$ is the Higgsino mass, ${\alpha(m_Z)}$ is the fine-structure constant, ${\theta_W}$ is the Weinberg angle, and $M_{\rm SUSY}$ is a representative supersymmetric mass scale.   This generic form of $a_{\mu}^{\rm SUSY} $ is proportional to $ m_{\mu}^2 / M_{\rm SUSY}^2$. The $1/M_{\rm SUSY}^2$-behavior of $a_{\mu}^{\rm SUSY}$ reflects the decoupling properties of SUSY, while its $m_{\mu}^2$-behavior reflects a chirality-flipping interaction between left-handed and right-handed muons. In detail, the Yukawa coupling for muons $Y_\mu$ breaks the chirality symmetry after electroweak symmetry breaking. The value of $Y_{\mu}$ is enhanced in the MSSM by a factor $1/\cos{\beta}\approx \tan{\beta}$ compared to its SM value, and this $Y_{\mu}$ enters the Feymann diagrams contributing to $a_{\mu}$ in the vertices where the muon chirality is flipped. Thus, supersymmetric particles in the mass range $100-500~{\rm GeV}$ could be the source of $\Delta a_{\mu}$.

\par The above-discussed favored mass range of supersymmetric particles in the SUSY interpretation of $\Delta a_{\mu}$  should be directly observable in the Large Hadron Collider (LHC). The corresponding parameter space, in general, predicts observable $3\ell + E_T^{\rm miss}$ signals via the electroweak channel $pp\to \widetilde{\chi}_i^0\widetilde{\chi}_k^{\pm}$ and $2\ell + E_{T}^{\rm miss}$ signals via the direct slepton search channel $pp \to \widetilde{\mu}^\pm \widetilde{\mu}^{\mp}$ and chargino pair production channel $pp \to \widetilde{\chi}_1^{\pm} \widetilde{\chi}_1^{\mp}$~\cite{Endo:2017zrj}.  However, for some explanation mechanism in MSSM, recent studies have demonstrated that this parameter space is seriously contracted during direct SUSY searches conducted using the A Toroidal LHC ApparatuS (ATLAS) and the compact muon spectrometer (CMS) at the LHC~\cite{Hagiwara:2017lse, Ibe:2019jbx, Tran:2018kxv}. A recent revisit study shows that the latest LHC constraints disfavour the MSSM parameter space smuons are lighter than charginos. So the MSSM explanation of $\Delta a_{\mu}$ indicates that $m_{\tilde{\mu}} \geq m_{\widetilde{\chi}_1^{\pm}}$\cite{Endo:2020mqz}. Accordingly, the two-loop contribution of $a_{\mu}^{\rm SUSY}$ is often taken into consideration to relax the parameter space to some extent. Moreover, recent efforts to directly detect dark matter (DM) and nucleon scattering have placed extreme limits on the parameter space~\cite{Beskidt:2012sk, Cao:2016cnv,Cao:2016nix,Cao:2018rix, Barbieri:2000gf, Giudice:2006sn, Yanagida:2017dao, Wang:2017vxj, Endo:2011xq, Abe:2002eq, Gogoladze:2015jua,Ajaib:2015ika, Ajaib:2014ana, Babu:2014lwa, Gogoladze:2014cha, Okada:2013ija, Kowalska:2015zja, Mohanty:2013soa, Akula:2013ioa, Ibe:2013oha, Wang:2015rli, Wang:2015nra}.
 In addition, \texttt{MasterCode} has been recently applied for obtaining the global fitting results of the eleven-parameter MSSM (pMSSM11) to constraints derived from LHC~13 TeV data and recent searches for DM scattering, and the results have demonstrated relatively strong constraints on the MSSM sub-TeV parameter space, where the results obtaining the best fit represented a compressed mass spectrum of $m_{\widetilde{B}} \approx m_{\widetilde{W}} \approx m_{\widetilde{\mu}} \sim 300~{\rm GeV}$  with the value of $a_{\mu}^{\rm SUSY}$ derived mainly from the Bino-Wino-smuon loop~\cite{Bagnaschi:2017tru}. As such, these experimental limitations detract from efforts to identify the sources of $a_{\mu}$ based on the MSSM.

\par Fortunately, other experimentally feasible sources of $a_{\mu}$ can be postulated. A good candidate can be based on chirality flipping or seesaw mechanisms, where the Yukawa coupling $Y_\nu$ of the Higgs field to a right-handed neutrino and a left-handed muon could also be a source of $a_{\mu}$, provided the adopted theory accommodates a right-handed neutrino. Among the various seesaw mechanism models, the inverse seesaw extended Next-to-Minimal Supersymmetric SM (ISS-NMSSM)~\cite{Deppisch:2004fa,Arina:2008bb,Abada:2010ym,delAguila:2019mvp} has generated particular interest. In contrast to the MSSM or standard NMSSM, the lightest sneutrino represents a promising DM candidate in the ISS-NMSSM~\cite{Cao:2017cjf}. The $\hat{S} \hat{\nu} \hat{X}$ term in the superpotential of the ISS-NMSSM ensures that the singlet Higgs superfield $\hat{S}$ not only plays a role in the Higgs sector, as it does in the standard NMSSM, but also plays key roles in the neutrino sector and DM annihilation in the early universe. In addition, the DM-nucleus scattering cross sections are naturally suppressed due to the singlet nature of the superfields $\hat{S}$ and $\hat{\nu}$, and thereby survive the harsh DM direct detection exclusion limit. As for the most sensitive channels at the LHC that enable direct search from the perspective of the ISS-NMSSM, we note that neutralinos mostly decay into a neutrino and a sneutrino, which is invisible to the detector. Meanwhile, charginos decay into a lepton and a sneutrino. Therefore, the most sensitive channel at the LHC for the ISS-NMSSM pertains to the $2\ell+E_{T}^{\rm miss}$ signals, not the $3\ell + E_{T}^{\rm miss}$ signals, as is the case for the standard NMSSM. As a result, the ISS-NMSSM can be expected to contribute profoundly to an explanation of $\Delta a_{\mu}$ because the value of $Y_\nu$ in the model can reach $\mathcal{O}(0.1)$, which is of the same order as gauge coupling $g_1$ and $g_2$. Therefore, the additional contributions due to chirality flipping by a sneutrino mass term and $Y_\nu$ can induce a sufficiently large $a_{\mu}^{\rm SUSY}$. However, to our best knowledge, this source of $a_{\mu}$ has not been investigated in the past.

\par The present study addresses this issue by applying the ISS-NMSSM toward explaining the anomalous magnetic moment of muons. This is a particularly pertinent issue at the present moment owing to the upcoming Fermilab E989 experiment. The remainder of this paper is organized as follows. The ISS-NMSSM and the properties of the corresponding contribution of $a_{\mu}^{\rm SUSY}$ are introduced in Sec.~{\ref{sec:model}}. In Sec.~{\ref{sec:na}}, we optimally scan the parameter space of the ISS-NMSSM according to various experimental constraints, and numerical analysis is applied to an optimal parameter space for assessing the potential of the ISS-NMSSM to contribute a sufficiently large value of $a_\mu^{\rm SUSY}$ to account for $\Delta a_{\mu}$. Finally, a brief summary is provided in Sec.~{\ref{sec:sum}}.

\section{\label{sec:model}Inverse seesaw mechanism extended NMSSM and the muon \texorpdfstring{$g-2$}{}}
\subsection{Sneutrino sector of the ISS-NMSSM}
Neutrino masses and mixings can be generated within the NMSSM framework via various standard seesaw mechanisms. In this work, the inverse seesaw mechanism is implemented within the NMSSM by adding two gauge singlet superfields $\hat{\nu}$ and $\hat{X}$ with opposite lepton numbers $L=-1$ and $L=+1$ respectively for each generation. Any $\Delta L = 1$ term in the superpotential is assumed to be forbidden. With the assumption of $R$-parity conservation, the ISS-NMSSM superpotential is given as follows~\cite{Abada:2010ym}.
\begin{equation}\label{eq:sp}
\begin{split}
    W &= Y_u\hat{Q} \cdot \hat{H}_u \hat{u} + Y_d \hat{H}_d \cdot \hat{Q} \hat{d} + Y_e \hat{H}_d \cdot \hat{L} \hat{e} + \lambda \hat{S} \hat{H}_u\cdot\hat{H}_d + \frac{\kappa}{3}\hat{S}^3\\
     &+ \frac{1}{2} \mu_X \hat{X}\hat{X} + \lambda_N \hat{S} \hat{\nu} \hat{X} + Y_{\nu} \hat{L}\cdot\hat{H}_u \hat{\nu}
    \end{split}
\end{equation}
Here, $\hat{H}_u$, $\hat{H}_d$, and $\hat{S}$ are Higgs superfields, $\hat{Q}$ and $\hat{L}$ respectively denote the SU(2) doublet quark and lepton superfields, and $\hat{u}$ and $\hat{d}$, and $\hat{e}$ are the SU(2) singlet up-type and down-type quark superfields, and charged lepton superfields, respectively. In addition, the Yukawa couplings for quarks and leptons are given as $Y_i$, where $i=u,d,e,\nu$, the term $\lambda_N$ represents Yukawa coupling  of singlet Higgs $\hat{S}$ to neutrinos $\hat{\nu}$ and $\hat{X}$, $\mu_X$ is an $X$ type neutrino mass term, respectively, and $\lambda$ and $\kappa$ are the Higgs interaction couplings. The only dimensional parameter $\mu_X$ is introduced for providing a $\Delta L=2$ term in the inverse seesaw mechanism, which also violates the $\mathbb{Z}_3$ symmetry of the superpotential. Generally, a tiny neutrino mass is obtained by treating $\mu_X$ as an extremely tiny effective mass parameter generated by some unknown dynamics.
It may be noted that the first line in Eq.~(\ref{eq:sp}) is the standard NMSSM superpotential with $\mathbb{Z}_3$ symmetry. The soft breaking terms are given as follows~\cite{Cao:2017cjf}.
\begin{equation}\label{eq:sbt}
\begin{split}
       V_{\rm soft} &= V_{\rm NMSSM} \\
       &+ m_{\nu}^2 \tilde{\nu}_R \tilde{\nu}_R^* + m_x^2 \tilde{x}\tilde{x}^* \\
       &+ \left(\frac{1}{2}B_{\mu_x} \tilde{x}\tilde{x} + \lambda_N A_{\lambda_N} S \tilde{\nu}_R^* \tilde{x} + Y_{\nu}A_{Y_{\nu}} \tilde{\nu}_R^* \tilde{L}\cdot H_u + {\rm h.c.}\right)
\end{split}
\end{equation}
Here, $V_{\rm NMSSM}$ is the NMSSM soft breaking term, $\tilde{\nu}_R$ and $\tilde{x}$ are the scalar components of $\hat{\nu}$ and $\hat{X}$, respectively, while all other fields are defined as they are in the standard NMSSM, and the dimensional quantities $m_{\nu,x}^2$, $m_{x}^2$, $B_{\mu_x}$, $A_{\lambda_N}$, and $A_{Y_{\nu}}$ are the soft breaking parameters.

\par In $R$-parity conserved ISS-NMSSM, the lightest sneutrino $\tilde{\nu}_1$ may be a better DM candidate under the stringent DM direct detection experimental constraint than the lightest neutralino $\widetilde{\chi}_1^0$ candidate~\cite{Cao:2019qng}. After decomposing sneutrino fields $\tilde{\nu}_i$ according to charge conjugation parity (CP) symmetry into CP-even and CP-odd parts $\tilde{\nu}_i = \dfrac{1}{\sqrt{2}}(\phi_i + i \sigma_i)$, the symmetric $9\times9$ mass matrix for the CP-odd sneutrinos can be given as
\begin{equation}\label{eq:sneumass}
    M_{\tilde{\nu}^I}^2 = \begin{pmatrix}
                        m_{\sigma_L \sigma_L} & m_{\sigma_R \sigma_L} & m_{\sigma_x \sigma_L}\\
                        m_{\sigma_L \sigma_R} & m_{\sigma_R \sigma_R} & m_{\sigma_x \sigma_R}\\
                        m_{\sigma_L \sigma_x} & m_{\sigma_R \sigma_x} & m_{\sigma_x \sigma_x}
                        \end{pmatrix},
\end{equation}
where the following definitions are applied in the basis of $(\sigma_L, \sigma_R, \sigma_x)$ with the terms $v_u$, $v_d$, and $v_s$ representing the vacuum expectation value (vev) of the Higgs fields $H_u$, $H_d$, and $S$, respectively.
\begin{small}
\begin{equation}\label{eq:sneumasselem}
    \begin{split}
        m_{\sigma_L \sigma_L} &= \frac{1}{4}\left[2 v_u^2 \Re(Y_\nu^T Y_\nu^*) + 4\Re(m_l^2) \right]  + \frac{1}{8}(g_1^2 + g_2^2) (-v_u^2 + v_d^2)\\
        m_{\sigma_L \sigma_R} &= -\frac{1}{2}v_d v_s \Re(\lambda Y_\nu^*) +\frac{1}{\sqrt{2}} v_u \Re(Y_\nu A_{Y_\nu})\\
        m_{\sigma_L \sigma_x} &= \frac{1}{2} v_s v_u \Re(\lambda_{N}^T Y_\nu^*)\\
        m_{\sigma_R \sigma_R} &= \frac{1}{4}\left[ 2 v_s^2 \Re(\lambda_N \lambda_N^\dag) + 2 v_u^2 \Re(Y_\nu Y_\nu^\dag) + 4 \Re(m_\nu^2) \right]\\
        m_{\sigma_R \sigma_x} &= \frac{1}{4} \left[ - v_d v_u (\lambda^* \lambda_N^T +\lambda\lambda_N^\dag) +v_s^2(\kappa\lambda_N^\dag + \kappa^* \lambda_N^T) + 2\sqrt{2} v_s \left( \Re(A_{\lambda_N}^T \lambda_N^T) -\Re(\mu_X \lambda_N^\dag)\right) \right]\\
        m_{\sigma_x \sigma_x} &= \frac{1}{2} v_s^2 \Re(\lambda_N^T \lambda_N^*) -\Re(B_{\mu_X}) + \Re(\mu_X \mu_X^*) + \Re(m_x^2)
    \end{split}
\end{equation}
\end{small}
This CP-odd sneutrino mass matrix $M_{\tilde{\nu}^I}^2$ is diagonalized as  $M_{\tilde{\nu}^I}^{2, \rm diag} = Z^I M_{\tilde{\nu}^I}^2 Z^{I, \dag}$ using a unitary rotation matrix $Z^I$. The CP-even sneutrino mass matrix $M_{\tilde{\nu}^R}^2$ and its unitary rotation matrix $Z^R$ can be obtained similarly, where  $M_{\tilde{\nu}^R}^2 = M_{\tilde{\nu}^L}^2 \mid_{\mu_{X}\to -\mu_{X}, B_{\mu_X}\to -B_{\mu_X}}$. An analysis of Eq.~(\ref{eq:sneumass}) and Eq.~(\ref{eq:sneumasselem}) indicates that the off-diagonal element $m_{\sigma_L \sigma_R}$ is proportional to $Y_{\nu}$, $m_{\sigma_R \sigma_x}$ is proportional to $\lambda_{N}$, and $m_{\sigma_L \sigma_x}$ is proportional to the product $Y_{\nu} \lambda_N$.
\par It is also noted that $Y_\nu$ and $\lambda_{N}$ play important roles in the ISS-NMSSM for determining neutrino mass as well. Experimental data based on the observation of active neutrino oscillation place a constraint on the unitary of the neutrino mass rotation matrix. This unitary constraint can be translated into a constraint on the input parameters~\citep{Cao:2019aam}, as follows:
	\begin{equation}\label{eq:uc}
		\frac{\lambda_{N_{e}} \mu}{Y_{\nu_e} \lambda v_u} > 14.1, \quad
		\frac{\lambda_{N_\mu} \mu}{Y_{\nu_{\mu}} \lambda v_u} > 33.7, \quad
		\frac{\lambda_{N_\tau}\mu}{Y_{\nu_\tau} \lambda v_u} > 9.4.
	\end{equation}
These inequalities indicate that, for given Higgs sector parameters $\lambda$, $\tan{\beta}$, and $\mu$, the coupling term $Y_{\nu}$ sets a lower limit for $\lambda_N$. The following discussion demonstrates that this unitary constraint greatly suppresses the value of $a_{\mu}^{\rm SUSY}$.

\subsection{\label{sec:22} Muon \texorpdfstring{$g-2$}{} in the ISS-NMSSM}
\begin{figure}[ht]
\centering
\includegraphics[]{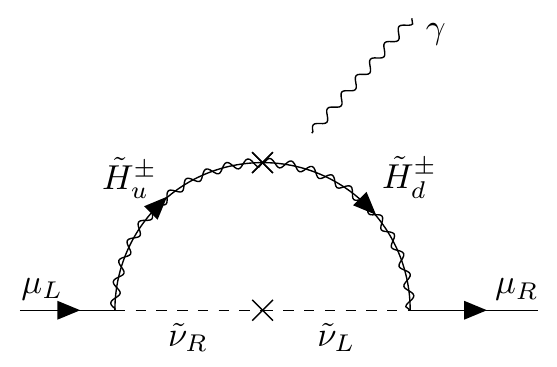}
\caption{\label{fig:feynloop}One-loop diagram of the Higgsino-sneutrino contribution to $a_{\mu}^{\rm SUSY}$, which is an additional contribution in the ISS-NMSSM compared with the MSSM.}
\end{figure}

\par As discussed in Section~\ref{sec:intro}, the muon magnetic moment $a_{\mu}$ always corresponds to a chirality-flipping interaction. Hence, the $\mu$ chirality must be flipped by one of the chirality-flipping interactions in each Feynman diagram contributing to $a_{\mu}$. The main chirality flipping interactions in the MSSM or standard NMSSM derive from the $\mu$-lepton line through the muon mass term, the Yukawa interaction of $H_d$ with $\mu_R$ and $\mu_L$ or $\nu_{\mu}$, the smuon line through the mass term $m_{\tilde{\mu}_L \tilde{\mu}_R}^2$, and SUSY Yukawa coupling of a Higgsino to a muon and $\tilde{\mu}$ or $\tilde{\nu}_\mu$. As indicated by Eq.~(\ref{eq:sp}), the ISS-NMSSM provides additional $\mu$ chirality flipping interactions from the muon-type sneutrino $\tilde{\nu}_\mu$ line through the mass term $m_{\tilde{\nu}_L \tilde{\nu}_R}$ and the additional SUSY Yukawa coupling of a Higgsino to $\tilde{\nu}_{\mu, R}$ and $\mu_L$ or $\nu_{\mu}$.

\par The above discussion indicates that only muon-type sneutrinos are related to $a_{\mu}$ in the ISS-NMSSM. Therefore, we can assume that no flavor mixing occurs in the sneutrino field and in the Yukawa couplings $Y_e$ and $Y_\nu$, and consider only the second generation according to the following notation: $y_\mu = Y_{e, 22}$, $y_\nu = Y_{\nu, 22}$, and substitute $3\times3$ matrices of muon-type sneutrino masses $m_{\tilde{\nu}^I}$ and $m_{\tilde{\nu}^R}$ and their corresponding $3\times3$ rotation matrices $Z^I$ and $Z^R$ to replace the $9\times9$ versions in the basis $(\tilde{\nu}_{\mu,L}, \tilde{\nu}_{\mu, R}, \tilde{\nu}_{\mu, x})$.
Therefore, the SUSY one-loop contribution to $a_{\mu}$ in the ISS-NMSSM is given as follows.
\begin{small}
\begin{equation}\label{eq:damufull}
\begin{split}
    &a_{\mu}^{\rm ISS-NMSSM} = a_{\mu}^{\widetilde{\chi}^0 \tilde{\mu}} + a_{\mu}^{\widetilde{\chi}^{\pm} \tilde{\nu}^I} + a_{\mu}^{\widetilde{\chi}^{\pm} \tilde{\nu}^R}\\
    a_{\mu}^{\widetilde{\chi}^0 \widetilde{\mu}} &= \frac{m_{\mu}}{16 \pi^2}\sum_{i,l}\left\{
    -\frac{m_{\mu}}{12 m_{\tilde{\mu}_l}^2} \left( |n_{il}^{\rm L}|^2 + |n_{il}^{\rm R}|^2 \right) F_1^{\rm N}(x_{il}) + \frac{m_{\widetilde{\chi}_i^0}}{3 m_{\widetilde{\mu}_l}^2} {\rm Re}(n_{il}^{\rm L} n_{il}^{\rm R}) F_2^{\rm N}(x_{il})
    \right\}\\
    a_{\mu}^{\widetilde{\chi}^\pm \tilde{\nu}^I} &= \frac{m_{\mu}}{16 \pi^2}\sum_{j,m}\left\{
    \frac{m_{\mu}}{12 m_{\tilde{\nu}_{\mu,m}^I}^2} \left( |c_{jm}^{I, \rm L}|^2 + |c_{jm}^{I, \rm R}|^2 \right) F_1^{\rm C}(x_{jm}) + \frac{2 m_{\widetilde{\chi}_j^\pm}}{3 m_{\tilde{\nu}_{\mu,m}^I}^2} {\rm Re}(c_{jm}^{I, \rm L}c_{jm}^{I, \rm R}) F_2^{\rm C}(x_{jm})
    \right\}\\
    a_{\mu}^{\widetilde{\chi}^\pm \tilde{\nu}^R} &= \frac{m_{\mu}}{16 \pi^2}\sum_{j,n}\left\{
    \frac{m_{\mu}}{12 m_{\tilde{\nu}_{\mu,n}^R}^2} \left( |c_{jn}^{R, \rm L}|^2 + |c_{jn}^{R, \rm R}|^2 \right) F_1^{\rm C}(x_{jn}) + \frac{2 m_{\widetilde{\chi}_j^\pm}}{3 m_{\tilde{\nu}_{\mu,n}^R}^2} {\rm Re}(c_{jn}^{R, \rm L}c_{jn}^{R, \rm R}) F_2^{\rm C}(x_{jn})
    \right\}
\end{split}
\end{equation}
\end{small}
Here, $i=1,\cdots,5$ and $j=1,2$ respectively denote the neutralino and chargino indices, $l=1,2$ denotes the smuon index, $m=1,2,3$ and $n=1,2,3$ denote the CP-odd and CP-even sneutrino indices, respectively, and
\begin{equation}
    \begin{split}
        n_{il}^{\rm L} = \frac{1}{\sqrt{2}}\left( g_2 N_{i2} + g_1 N_{i1} \right)X^*_{l1} -y_{\mu} N_{i3}X^*_{l2}, \quad
        &n_{il}^{\rm R} = \sqrt{2} g_1 N_{i1} X_{l2} + y_{\mu} N_{i3} X_{l1},\\
        c_{jm}^{I, \rm L} =\frac{1}{\sqrt{2}} \left( -g_2 V_{j1} Z_{m1}^{I,*} + y_{\nu} V_{j2} Z_{m2}^{I,*} \right), \quad
        &c_{jm}^{I, \rm R} =\frac{1}{\sqrt{2}} y_{\mu} U_{j2} Z_{m1}^I,\\
        c_{jn}^{R, \rm L} = \frac{1}{\sqrt{2}} \left( -g_2 V_{j1} Z_{n1}^{R,*} + y_{\nu} V_{j2} Z_{n2}^{R,*}\right), \quad
        &c_{jn}^{R, \rm R} = \frac{1}{\sqrt{2}} y_{\mu} U_{j2} Z_{n1}^R.
    \end{split}
\end{equation}
Here, $N$ is the neutralino mass rotation matrix, $X$ is the smuon mass rotation matrix, and $U$ and $V$ are the chargino mass rotation matrices defined by $U^* \mathcal{M}_{\widetilde{\chi}^\pm} V^\dag = m_{\widetilde{\chi}^\pm}^{\rm diag}$, where $\mathcal{M}_{\widetilde{\chi}^\pm}$ is the chargino mass matrix.
The kinematic loop functions depend on the variables $x_{il}=m_{\widetilde{\chi}_i^0}^2 / m_{\tilde{\mu}_l}^2$, $x_{jm} = m_{\widetilde{\chi}_j^\pm}^2 / m_{\tilde{\nu}_{\mu, m}^I}^2$, and $x_{jn} = m_{\widetilde{\chi}_j^{\pm}}^2 / m_{\tilde{\nu}_{\mu, n}^R}^2$, and are given as follows.
\begin{equation}
    \begin{split}
        F_1^{\rm N}(x) &= \frac{2}{(1-x)^4} \left( 1 - 6x + 3x^2 + 2x^3 - 6x^2 \ln{x} \right)\\
        F_2^{\rm N}(x) &= \frac{3}{(1-x)^3} \left( 1 - x^2 + 2x \ln{x} \right)\\
        F_1^{\rm C}(x) &= \frac{2}{(1-x)^4} \left( 2 + 3x - 6x^2 + x^3 + 6x \ln{x} \right) \\
        F_2^{\rm C}(x) &= -\frac{3}{2(1-x)^3} \left( 3 - 4x + x^2 + 2\ln{x} \right)
    \end{split}
\end{equation}
All of the above $F(x)$ functions are normalized with conditions $F_i^j(1) = 1$, where $x=1$ correspond to degenerate sparticles.

\begin{table}[ht]
\centering
\resizebox{0.95\textwidth}{!}{
\begin{tabular}{cccccccc|c}
\hline
$\lambda$ & $\kappa$   & $\tan{\beta}$ & $\mu$    & $M_1$    & $M_2$     & $m_{\ell_\mu}$ & $m_{E_\mu}$   & \multirow{2}{*}{$a_{\mu}^{\rm SUSY}$}   \\
0.1    & 0.6     & 50      & 350 GeV     & 3000 GeV  & 3000 GeV  & 500 GeV  & 3000 GeV &                           \\ \cline{1-8}
$y_{\nu}$    & $\lambda_{N_\mu}$ & $A_{y_{\nu}}$    & $A_{\lambda_{N_\mu}}$ & $m_{\nu}^2$  & $m_x^2$    & $\mu_X$  & $B_{\mu_X}$ & \multirow{2}{*}{$7.46\times10^{-10}$} \\
0.2    & 0.3     & -1000 GeV   & -3000 GeV   & $(200~{\rm GeV})^2$ & $(800~{\rm GeV})^2$ & 0    & 0    &                           \\ \hline
\end{tabular}}
\caption{\label{tab:bps} Benchmark parameter settings of the heavy Bino, Wino, and right-handed smuon limits, and the corresponding value of $a_{\mu}^{\rm SUSY}$ obtained from Eq.~(\ref{eq:simau}). In this case, the dominant contribution to $a_{\mu}^{\rm SUSY}$ derives from the Higgsino-sneutrino. }
\end{table}
\par At the heavy Bino, Wino, and right-handed smuon limits, the contribution of the Higgsino-sneutrino (HS) loop shown in Fig.~\ref{fig:feynloop} to $a_{\mu}^{\rm SUSY}$ is dominant, where the $\mu$-chirality flipping derives from the left-right handed sneutrino transition term $m_{\tilde{\nu}_L \tilde{\nu}_R}^2$ in the sneutrino mass matrix. Accordingly, $a_{\mu}^{\rm SUSY}$ can be expressed as
\begin{equation}\label{eq:simau}
	a_{\mu}^{\rm SUSY} \approx \frac{m_{\mu} y_\mu y_\nu}{48 \pi^2} \mu \left\{ \sum_{m} \frac{Z^{I, *}_{m2} Z^{I}_{m1}}{m_{\tilde{\nu}_{\mu, m}^I}^2} F_2^C(x_{1m}) + \sum_{n} \frac{Z^{R, *}_{n2} Z^{R}_{n1}}{m_{\tilde{\nu}_{\mu, m}^R}^2} F_2^C(x_{1n})  \right\} + \mathcal{O}\left( \frac{\mu}{M_2} \right).
\end{equation}
\begin{figure}[ht]
\centering
	\subfigure[] {\includegraphics[width=0.32\textwidth]{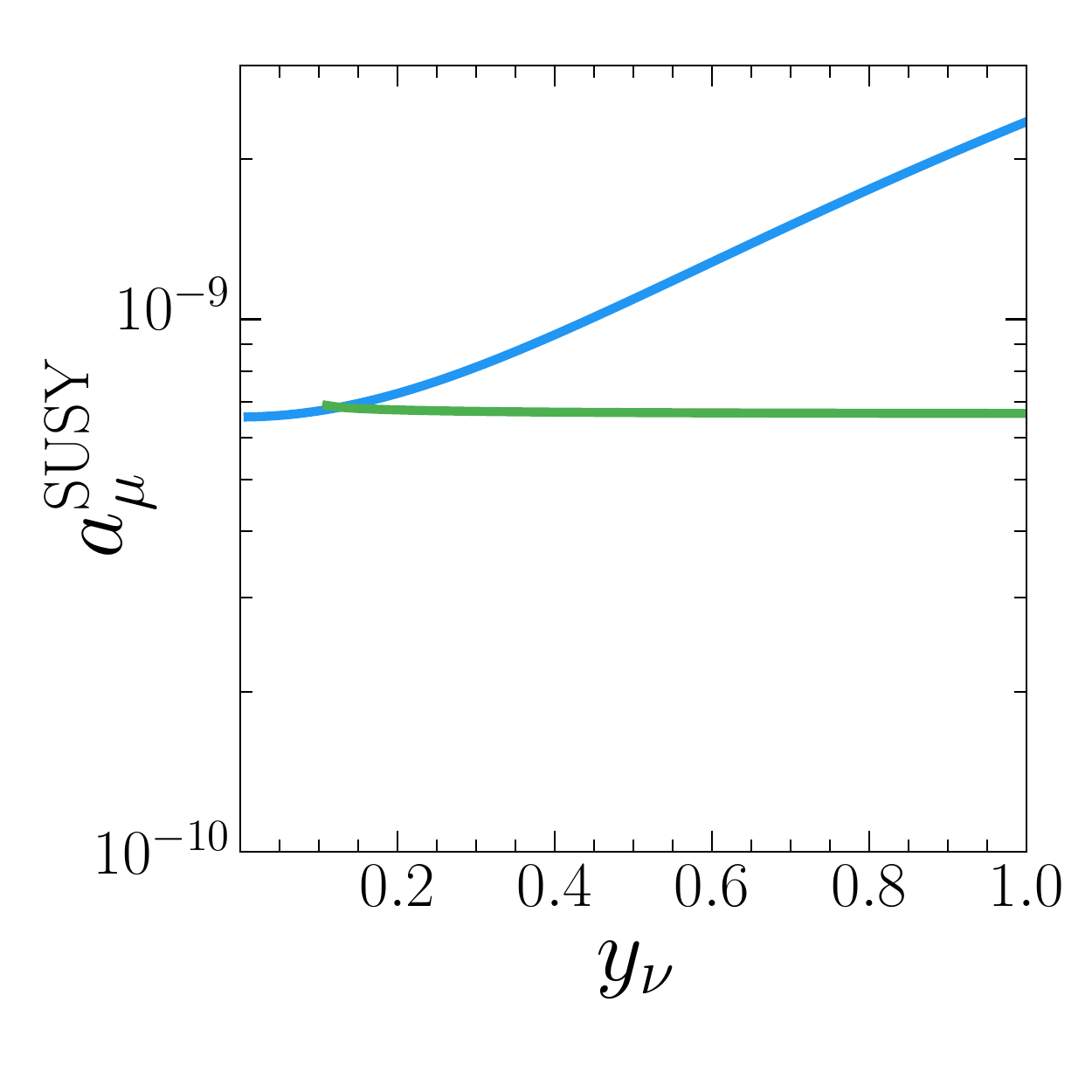}}
	\subfigure[] {\includegraphics[width=0.32\textwidth]{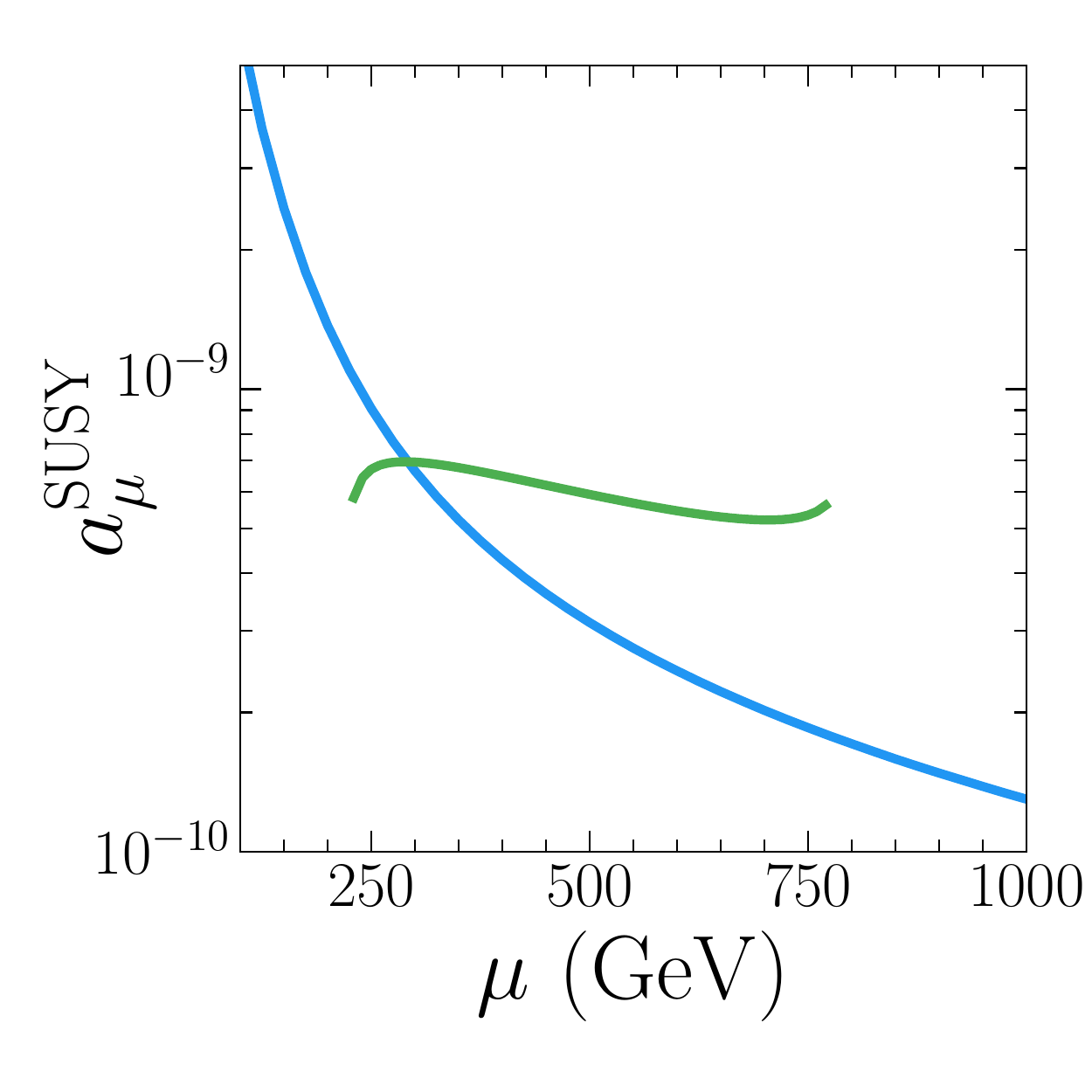}}
	\subfigure[] {\includegraphics[width=0.32\textwidth]{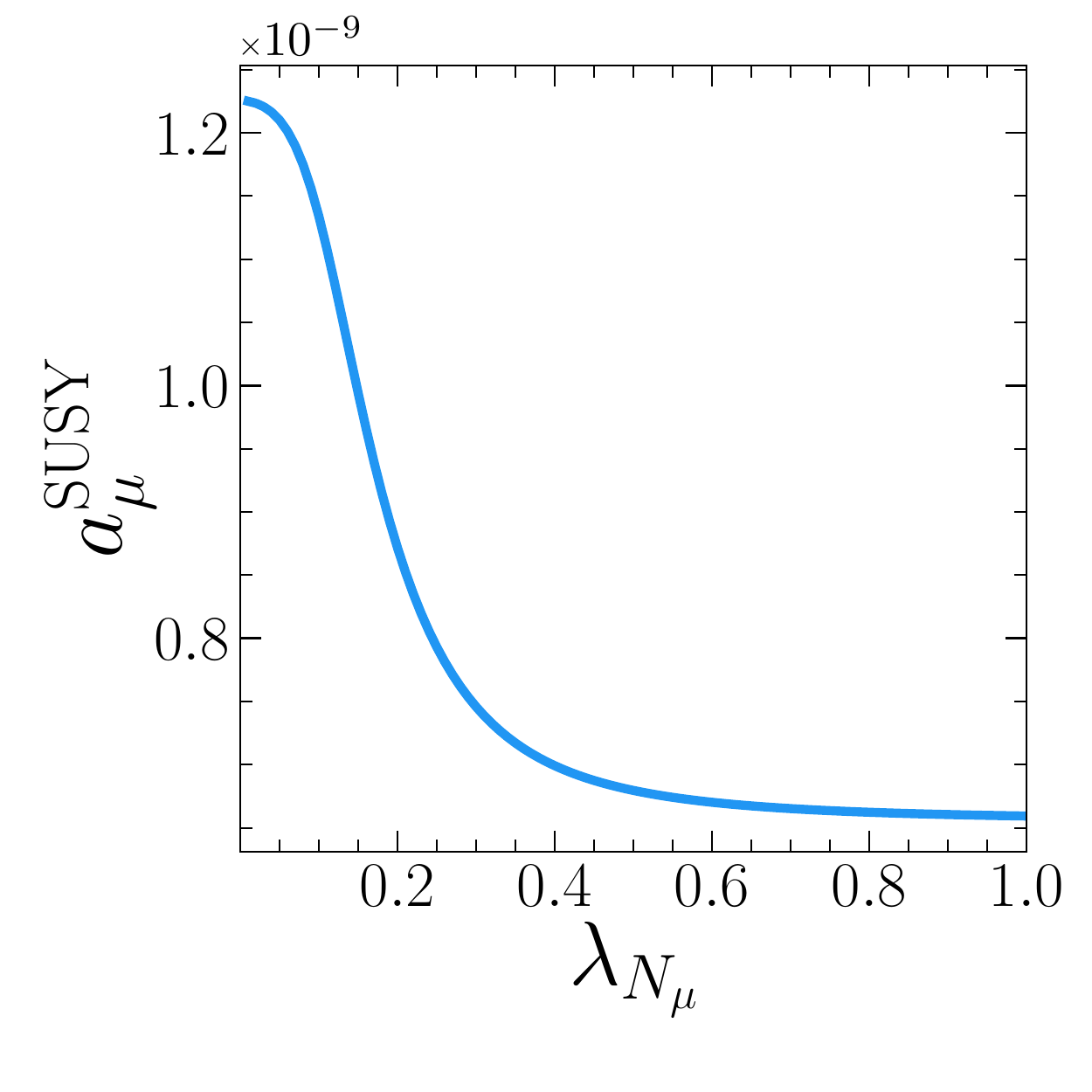}}\\
	\subfigure[] {\includegraphics[width=0.32\textwidth]{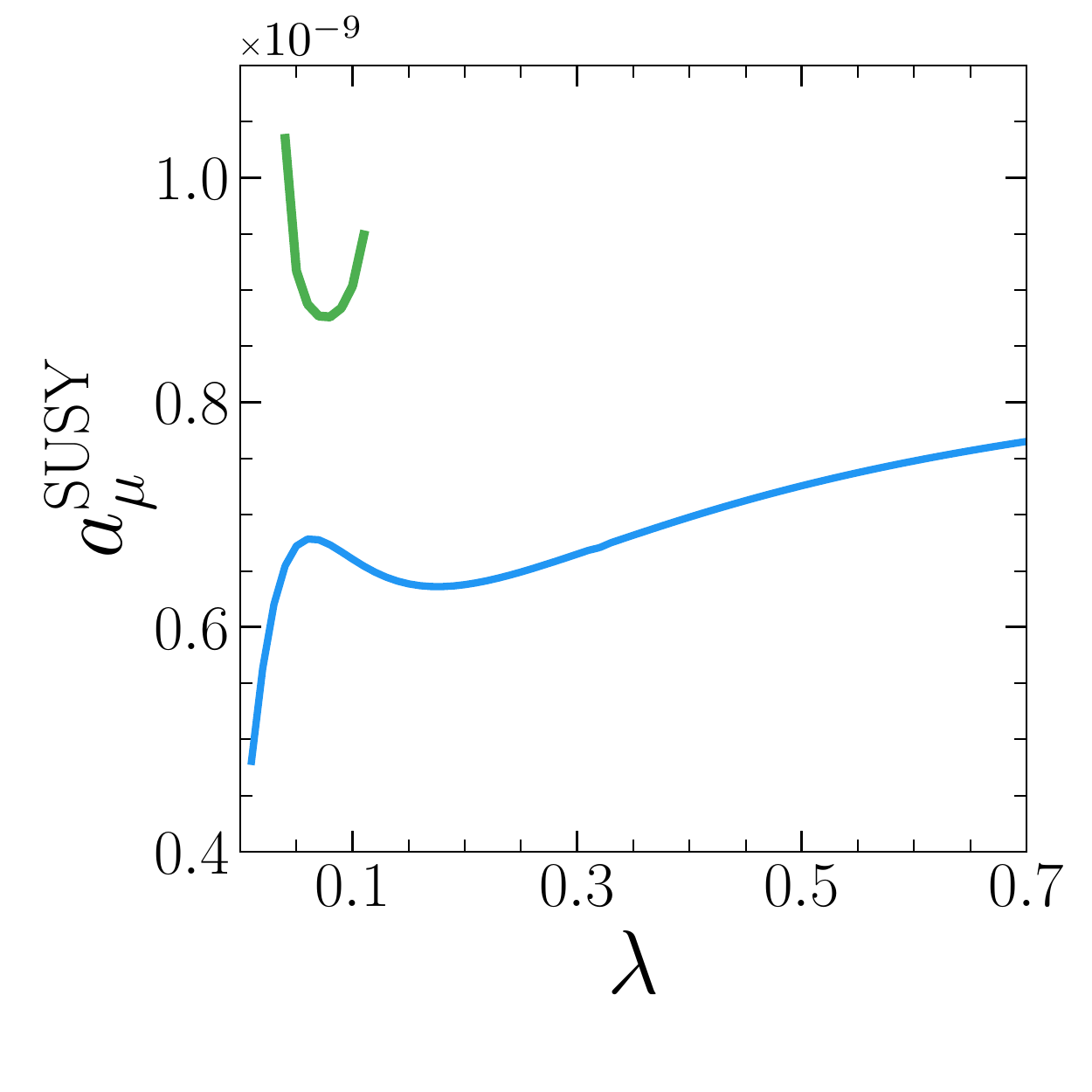}}
	\subfigure[] {\includegraphics[width=0.32\textwidth]{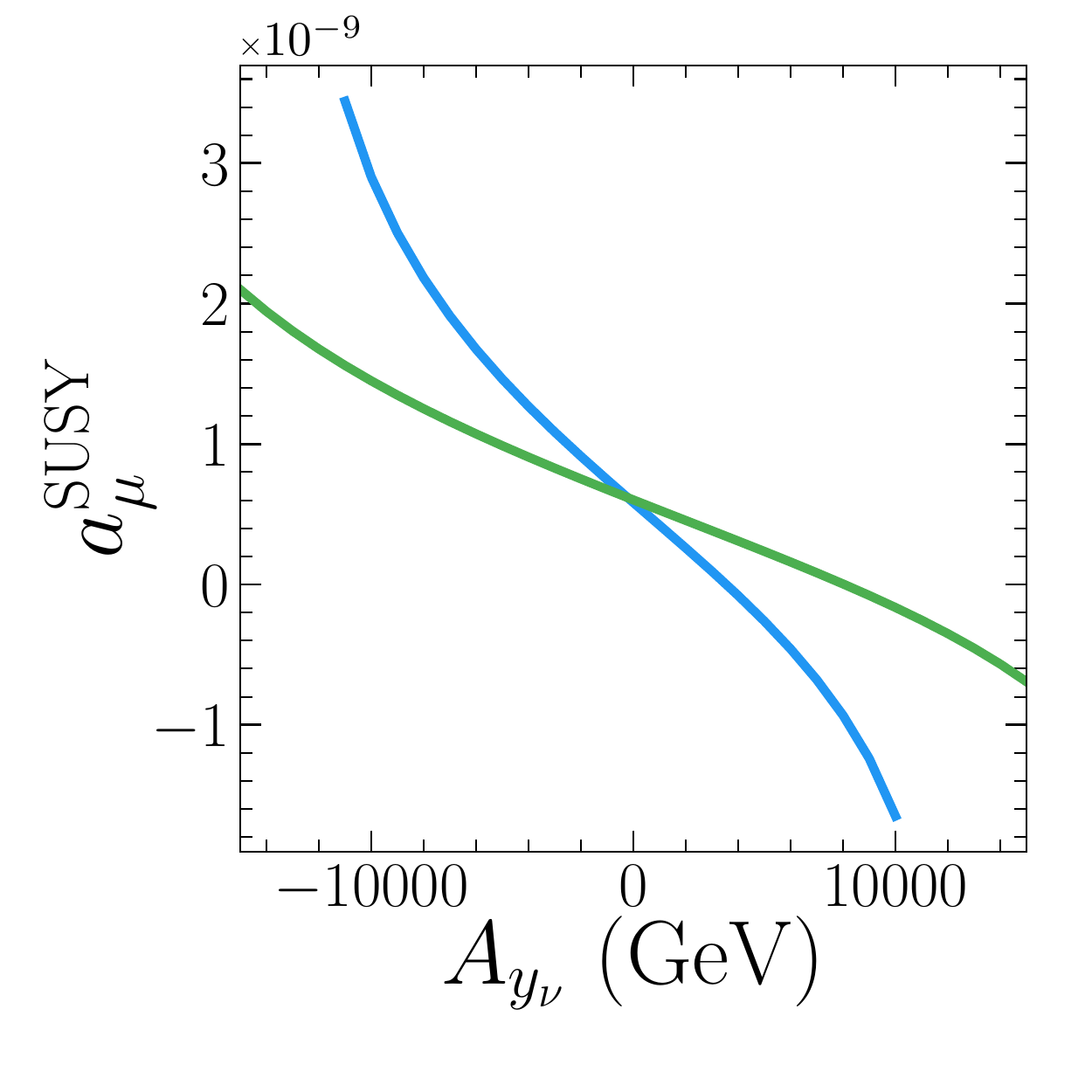}}
	\subfigure[] {\includegraphics[width=0.32\textwidth]{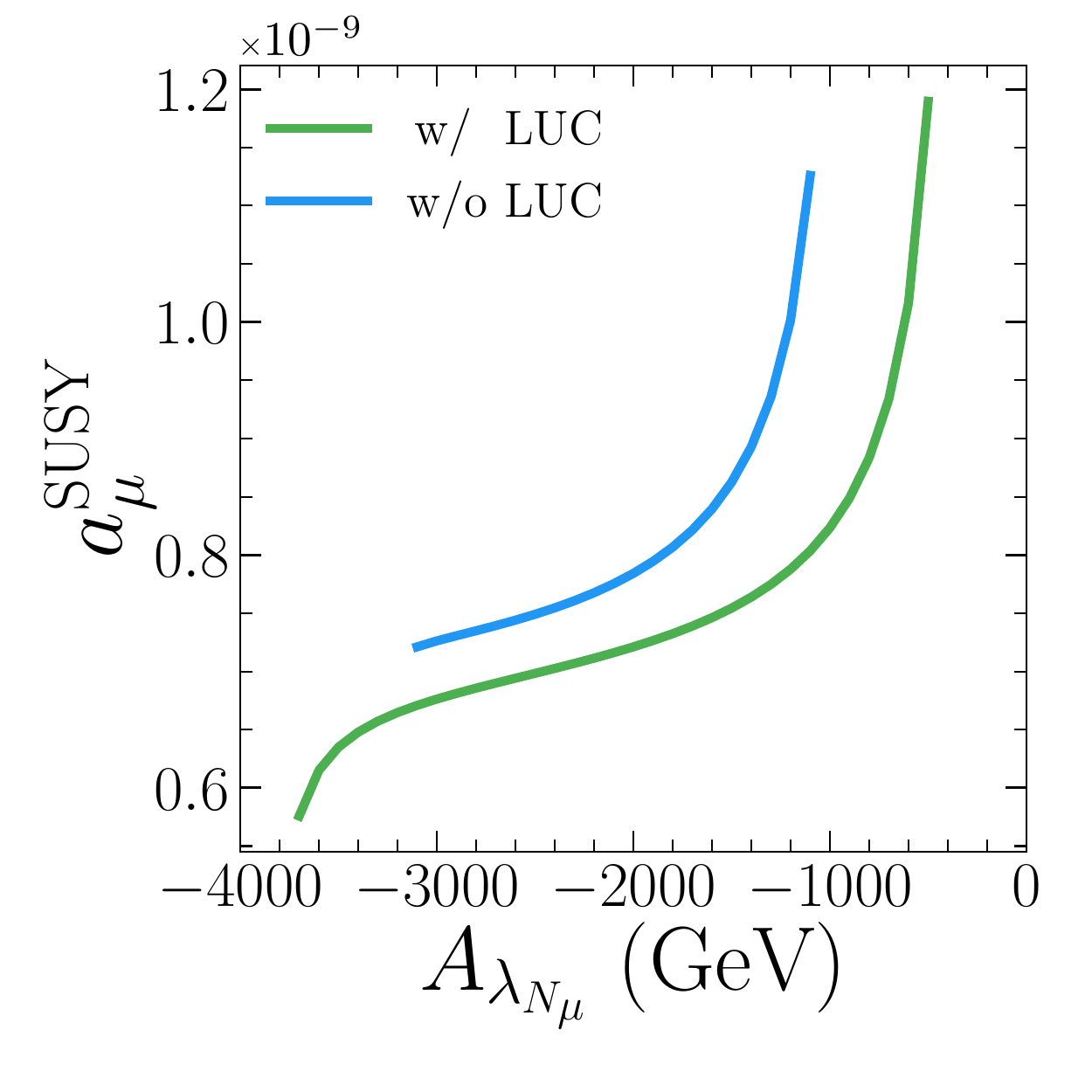}}
\caption{\label{fig:para} Values of $a_{\mu}^{\rm SUSY}$ obtained from Eq.~(\ref{eq:simau}) as functions of select parameters in the ISS-NMSSM with all other parameters obtained from Table~\ref{tab:bps}. The green lines in the plots were obtained when taking the leptonic unitary condition (LUC) into account by setting the value of $\lambda_{N_\mu} = (33.7 y_\nu \lambda v_u)/\mu$, while the blue lines do not consider the LUC, and simply employ the benchmark value of $\lambda_N$ given in Table~\ref{tab:bps}.}
\end{figure}
However, the HS contribution to $a_{\mu}^{\rm SUSY}$ is difficult to estimate due to the complexity of the $3\times3$ sneutrino mass matrix. We demonstrate this complexity and the influence of the theoretical input parameters on $a_{\mu}^{\rm SUSY}$ by plotting the values of $a_{\mu}^{\rm SUSY}$ obtained from Eq.~(\ref{eq:simau}) as functions of select parameters in the ISS-NMSSM with all other parameters obtained from Table~\ref{tab:bps}. The results are shown in Fig.~\ref{fig:para}, where the blue lines in all plots represent the values of $a_{\mu}^{\rm SUSY}$ obtained as functions of the input parameters without considering the leptonic unitary constraint (LUC), while the green lines represent the cases where the LUC is taken into consideration by setting the value of $\lambda_{N_\mu}=(33.7 y_\nu \lambda v_u)/\mu$. An analysis of the results in Fig.~\ref{fig:para} yields the following observations.
\begin{itemize}
\item When the LUC is not taken into consideration, $a_{\mu}^{\rm SUSY}$ increases monotonically with increasing $y_{\nu}$ (Fig.~\ref{fig:para}(a)), as would be expected from Eq.~(\ref{eq:simau}), and decreases monotonically with respect to $\mu$ (Fig.~\ref{fig:para}(b)) due to the monotonically decreasing loop function $F_2^{C}$. In addition, $a_{\mu}^{\rm SUSY}$ decreases monotonically with increasing $\lambda_{N_{\mu}}$ (Fig.~\ref{fig:para}(c)) because $\lambda_{N_{\mu}}$ affects the mass of the right-handed sneutrino according to the $(\lambda_{N_{\mu}} v_s)^2$ term in the sneutrino mass matrix term $m_{\tilde{\nu}_R \tilde{\nu}_R}^2$, as shown in Eq.~(\ref{eq:sneumasselem}). However, the dependence of $a_{\mu}^{\rm SUSY}$ on $\lambda$ (Fig.~\ref{fig:para}(d)) is quite complicated because the Higgs self-coupling term $\lambda$ is inversely proportional to $v_s$ for a given $\mu$, which affects the sneutrino mass terms $m_{\tilde{\nu}_L \tilde{\nu}_x}^2$, $m_{\tilde{\nu}_R \tilde{\nu}_R}^2$, $m_{\tilde{\nu}_R \tilde{\nu}_x}^2$, and $m_{\tilde{\nu}_x \tilde{\nu}_x}^2$. Nonetheless, we note from the plot that $a_{\mu}^{\rm SUSY}$ is enhanced within a small range $\lambda$ around 0.1. \item If the sneutrinos dominated by the scalar $\tilde{x}$ field have sufficient mass, the sneutrino mixing matrix $Z$ roughly satisfies the relation $Z_{12}Z_{11} \approx -Z_{22}Z_{21}$. Therefore, the cancellation between two light sneutrino contributions to $a_{\mu}^{\rm SUSY}$ greatly suppresses $a_\mu$, and $a_{\mu}^{\rm SUSY}$ presents the following trend:
\begin{equation}\label{eq:canc}
 	a_{\mu}^{\rm SUSY} \propto |Z_{11}Z_{12}|\left(\frac{1}{m_{\tilde{\nu}_1}^2} - \frac{1}{m_{\tilde{\nu}_2}^2} \right).
\end{equation}

\item The soft breaking terms $A_{y_\nu}$ and $A_{\lambda_{N_{\mu}}}$ govern the mixing between the left-handed and right-handed sneutrinos and the scalar $\tilde{x}$ field, and thereby affect the squared mass of the lightest sneutrino. In addition, the signs of $A_{y_\nu}$ and $A_{\lambda_{N_\mu}}$ respectively affect the signs of the products $Z_{m2}Z_{m1}$ and $Z_{n1}Z_{n2}$ in Eq.~(\ref{eq:simau}). Therefore, the sign of $a_{\mu}^{\rm SUSY}$ can differ from the sign of $\mu$ or $M_2 \mu$ in the ISS-NMSSM, which is shown in Fig.~\ref{fig:para}(e). Moreover, overly large values of $A_{y_\nu}$ and $A_{\lambda_{N_{\mu}}}$ can result in a negative squared sneutrino mass, which is unphysical.
However, we note from Fig.~\ref{fig:para}(e) that $a_{\mu}^{\rm SUSY}$ can be increased up to $2\times10^{-9}$ with a sufficiently large value of $|A_{y_\nu}|$. This can be attributed to two reasons: one is that $|Z_{11}Z_{12}|$ is proportional to $A_{y_\nu}$, and the other is that the mass splitting between the left-handed sneutrino and the right-handed sneutrino also increases with increasing $|A_{y_\nu}|$. As shown in Fig.~\ref{fig:para}(f), mixing between the $\tilde{x}$ field and the right-handed sneutrino field has a non-trivial contribution to the value of $a_{\mu}^{\rm SUSY}$ because this mixing effect may induce the strong cancellation associated with Eq.~({\ref{eq:canc}}).

\item The significant differences between the green and blue lines in Fig.~{\ref{fig:para}} indicate that the LUC also has a complicated impact on $a_{\mu}^{\rm SUSY}$. We note from Figs.~{\ref{fig:para}}(a) and (c) that, in the absence of the LUC, the increase in $a_{\mu}^{\rm SUSY}$ with increasing $y_\nu$ would be diminished by an increasing $\lambda_{N_{\mu}}$. The green line in Fig.~{\ref{fig:para}}(a) is a revealing representation of the effect of the LUC on $a_{\mu}^{\rm SUSY}$. In addition, the behaviors of $a_{\mu}^{\rm SUSY}$ with respect to $\mu$ and $\lambda$ are significantly affected by the LUC due to the important role of the $\lambda_{N_{\mu}} v_s$ term in the sneutrino mass matrix. Moreover, the physical $\mu$ and $\lambda$ parameter spaces are significantly reduced under the LUC. However, $a_{\mu}^{\rm SUSY}$ can still attain values on the order of $10^{-9}$.
\end{itemize}

\par In summary, the HS contribution to $a_{\mu}^{\rm SUSY}$ is limited significantly by the cancellation effect between different sneutrino contributions and by the LUC. However, mixing between the $\tilde{x}$ field and the right-handed sneutrino field ensures an HS contribution that is sufficiently large to explain $\Delta a_{\mu}$.

\section{Numerical analysis}\label{sec:na}
In this section, we explore the HS to $a_{\mu}^{\rm SUSY}$ comprehensively. Due to the ``curse of dimensionality'', the solution to a complex high-dimensional and multimodal distribution often requires very time-consuming fitness function evaluations, substantial computing resources and the efficient parameter space scan technique. For example, in the study of the Higgs sector in the NMSSM, there are several scan techniques used in previous researches, e.g. random sampling~\cite{Baum:2017gbj} and \texttt{Minuit} fit technique~\cite{Beskidt:2019mos}. Previous discussion indicate that the parameters $\lambda$ and $\mu$ are sensitive to $a_{\mu}^{\rm SUSY}$ and affected by the unitary of the neutrino mass rotation matrix. In the ISS-NMSSM, these parameters also play essential roles in the DM physics, Higgs boson properties and the collider phenomenology. 
In this work, we first take the genetic algorithm (GA)~\cite{MCCALL2005205} technique to scan the 18 dimensional ISS-NMSSM parameter space to find a sample that satisfies all the constraints under the current circumstance. In such a high-dimensional parameter space computing, the advantage of GA technique is its good global searching capability, computing cheapness and the stability of the result. However, the optimal sample achieved by the GA computing can not reflect the overall predictions of $a_{\mu}^{\rm SUSY}$ and the characteristics of the theory. 
Concerning this shortcoming, we note that the sneutrino DM in the ISS-NMSSM with moderately large $\mu$ can easily predict the observed DM relic density and coincide with the current DM direct detection experiments and the Electroweakino searches at LHC~\cite{Cao:2019qng}. This inspires us to use the \texttt{MultiNest}~\cite{Feroz:2008xx, Feroz:2013hea} algorithm to scan only the parameter space related to the HS contribution. In this way, the shortcomings of the GA technique are overcame without affecting the generality of our results.  

\subsection{Analysis of benchmark sample}
In order to answer whether ISS-NMSSM can predict a relatively large HS contribution to $a_{\mu}^{\rm SUSY}$ without contradicting the current experimental observations. We take GA method to scan the ISS-NMSSM parameter space with the following settings:
\begin{equation}\label{eq:para}
\begin{split}
    &0 < \lambda < 0.7, \quad |\kappa| < 0.7, \quad 1 < \tan{\beta} < 60, \quad 100~{\rm GeV} < \mu < 600~{\rm GeV},\\
    &\left| A_{\kappa} \right| < 500~{\rm GeV}, \quad A_{\lambda} = 2~{\rm TeV}, \quad \left| A_{t} \right| < 5~{\rm TeV}, \quad A_{b} = A_{t},  \\
    &0 < Y_{\nu_\tau} < 0.5, \quad \left| A_{Y_{\nu_\tau}} \right| < 2~{\rm TeV},\quad 0 < \lambda_{N_\tau} < 0.5, \quad \left| A_{\lambda_{N_\tau}} \right| < 2~{\rm TeV},\\
    &m_{\ell_\tau} = 800~{\rm GeV}, \quad 0 < m_{\nu_{\tau}} < 500~{\rm GeV},\quad 0 < m_{x_{\tau}} < 500~{\rm GeV},\\
    &0 < y_{\nu} < 0.5, \quad \left| A_{y_{\nu}} \right| < 3~{\rm TeV}, \quad 0 < \lambda_{N_\mu} < 0.5, \quad \left| A_{\lambda_{N_\mu}} \right| < 3~{\rm TeV}, \\
    &100~{\rm GeV} < m_{\ell_\mu} < 500~{\rm GeV}, \quad \left| m_{\nu}^2 \right| < (500~{\rm GeV})^2, \quad  m_{x} = 800~{\rm GeV},
\end{split}
\end{equation}
with all the parameters defined at the scale of $1~{\rm TeV}$. All other parameters, like those related to the squark, first generation sparticle, and gauginos, are fixed at a common value of $3~{\rm TeV}$. The parameter settings given in Eq.~(\ref{eq:para}) include the following assumptions.
\begin{itemize}
	\item Contributions to $a_{\mu}$ from the standard NMSSM require that the Bino soft mass $M_1$, Wino soft mass $M_2$, and Higgsino mass $\mu$ must be $\mathcal{O}(100)~{\rm GeV}$.  As such, $M_1$ and $M_2$ are too large in the standard NMSSM to provide sufficient contributions to $a_\mu$. Therefore, the SUSY contribution to $a_{\mu}$ can only derive from the HS loop illustrated in Fig.~\ref{fig:feynloop}.
	\item As established in previous studies~\cite{Cao:2017cjf, Cao:2019qng}, mass splitting between CP-even and CP-odd sneutrinos is related to the parameters $\mu_x$ and $B_{\mu_x}$. Therefore, this mass splitting can be neglected in the discussion of $a_{\mu}$ by assuming that the masses and the rotation matrices of the CP-even and CP-odd sneutrinos are equivalent\footnote{In the case of $B_{\mu_x} = 0$, the sneutrinos are complex fields, and the  DM-nucleon scattering rate obtains an additional contribution from the $t$-channel via the $Z$ boson. }.
	\item The advantages of the DM properties of sneutrinos in the ISS-NMSSM are preserved by setting the third generation sneutrino parameters to provide a $\tau$-type sneutrino DM candidate, which avoids the restrictions associated with DM observations of the $\mu$-type sneutrino.
	\item In the right-handed $\tau$-type sneutrino DM case, Higgsino dominated neutralinos (chargino) decay into a sneutrino plus a neutrino ($\tau$ lepton), i.e., ${\rm Br}(\widetilde{\chi}_{1,2}^0\to \tilde{\nu} \nu_{\tau})={\rm Br}(\widetilde{\chi}_1^{\pm}\to \tilde{\nu} \tau^\pm)=1$. Therefore, the most sensitive channel at the LHC is the chargino pair direct search $pp \to \widetilde{\chi}_1^+ \widetilde{\chi}_1^- \to 2\tau + E_{\rm T}^{\rm miss}$ channel, where the detection limit for the lightest supersymmetric particle (LSP) is about $300~{\rm GeV}$~\cite{Aaboud:2307399}. Therefore, we assume that the lightest sneutrino $\tilde{\nu}_1$ is a $\tau$-type sneutrino with a mass greater than $300~{\rm GeV}$.
	\item Undoubtedly, due to the soft breaking term $V_{\rm soft}$ in Eq.~({\ref{eq:sbt}}), the additional introduced sneutrino fields are embedded into the neutral scalar field potential, which is relevant for electroweak symmetry breaking. As a result, the sneutrino fields can acquire non-zero vevs, which will lead to $R$-parity breaking, various mixings between leptons with charginos and neutralinos, and mixings between Higgs bosons and sleptons~\cite{Liu:2005rs}. Of significance here is that the LSP $\tilde{\nu}_1$ is unstable, and $\tilde{\nu}_1$ can decay into two leptons. However, this contradicts our previous assumptions.  Therefore, we assume that sneutrino fields cannot acquire non-zero vevs, and this would further limit the parameter space of the theory.
\end{itemize}
In the scanning calculations, the ISS-NMSSM model file is generated by the Mathematica package \texttt{SARAH}~\cite{Staub:2015kfa}, the particle spectrum and the value of $a_{\mu}^{\rm SUSY}$ are generated using the \texttt{SPheno} program~\cite{Porod:2003um, Porod:2011nf}, the DM relic density and DM direct detection cross sections are computed using the \texttt{micrOMEGAs}~\cite{Belanger:2010pz}
code, and electroweak vacuum stability and sneutrino stability are tested using the \texttt{Vevacious} program~\cite{Camargo-Molina:2013qva,Camargo-Molina:2014pwa}, where the tunneling time from the input electroweak minimum to the true minimum is estimated using the \texttt{CosmoTransitions} program~\cite{Wainwright:2011kj} if needed. The optimal parameter set is obtained by GA minimization based on the following $\chi^2$ function with the 18-dimension free parameter space given by Eq.~(\ref{eq:para}):
\begin{equation}
	\chi^2 = \chi^2_{\rm Higgs} + \chi^2_{B} + \chi^2_{\rm DM} + \chi^2_{\rm Unitary} + \chi^2_{a_{\mu}} + \chi^2_{\rm vev} + \chi^2_{\rm veto}.
\end{equation}
The individual $\chi^2$ terms in the above equation are defined as follows.
\begin{itemize}
	\item $\chi^2_{\rm Higgs} = \frac{( m_{h} - m_{h}^{\rm obs})^2}{2\sigma_{m_{h}}^2} + \chi^2_{\texttt{HB}} + \chi^2_{\texttt{HS}} $: Here,  $m_h$ is the theoretical prediction, $m_{h}^{\rm obs} = 125.18~{\rm GeV}$~\cite{Sirunyan:2272260,ATLAS-CONF-2017-046,Tanabashi:2018oca} is the measured value, $\sigma_{m_{h}} = 3~{\rm GeV}$ is the total (theoretical and experimental) uncertainty, $\chi^2_{\texttt{HB}} = 0$ if the sample satisfies constraints associated with the direct search for Higgs bosons at the Large Electron-Positron (LEP) collider and Tevatron collider based on calculations using the \texttt{HiggsBounds} code~\cite{Bechtle:2008jh,Bechtle:2011sb}, and $\chi^2_{\texttt{HS}} = 0$ if the SM-like Higgs boson in the sample is compatible with current experimental observations, which is tested using the \texttt{HiggsSignals} code~\cite{Bechtle:2013xfa, Bechtle:2014ewa}. Otherwise, $\chi^2_{\texttt{HB}}$ or $\chi^2_{\texttt{HS}}$ is equal to 10000.
	\item $\chi^2_{B} = \frac{1}{2}\sum_i \left(\frac{\mathcal{O}_{\rm th}^i - \mathcal{O}_{\rm obs}^i}{\sigma^i}\right)^2$: $B$-physics observations ${\rm BR}(B_s \to \mu^+ \mu^-)$ and ${\rm BR}(B_s \to X_s \gamma)$~\cite{Tanabashi:2018oca} are take into consideration in this work, and both introduce standard Gaussian constraints into $\chi^2$.
	\item $\chi^2_{\rm DM} = \frac{(\Omega h^2_{\rm th} - \Omega h^2_{\rm obs} )^2}{2\sigma_{\Omega h^2}^2} + \chi^2_{\rm DMDD}$: Here, $\Omega h^2_{\rm th}$ is the theoretical prediction of the DM relic density, $\Omega h^2_{\rm obs} = 0.120$ is the cosmological DM parameter obtained in the latest PLANCK report \cite{Aghanim:2018eyx}, and $\sigma_{\Omega h^2} = 0.0120$ is the total uncertainty. The term $\chi^2_{\rm DMDD} = 0$ if the DM-nucleon scattering cross section is less than the current $90\%$ upper limits established by the Xenon-1T 2018 report \cite{Aprile:2018dbl}; otherwise, $\chi^2_{\rm DMDD} = 10000$.
	\item The unitary constraints of the second and third generations are included in the $\chi^2_{\rm Unitary}$ term as follows.
	\begin{equation}
		\chi^2_{\rm Unitary} = \sum_{i=\mu, \tau} \chi^2_{{\rm Unitary}, i}, \quad
		\chi^2_{{\rm Unitary}, i} =
		\begin{cases}
			100(r_i - r_i^{\rm low})^2 ,	&r_i < r_i^{\rm low} \\
			0,	&r_i \geq r_i^{\rm low}
		\end{cases}
	\end{equation}
	Here, $r_i = (\lambda_{N_i} \mu)/(Y_{\nu_i} \lambda v_u)$, $r_\mu^{\rm low} = 33.7$, and $r_{\tau}^{\rm low} = 9.4$.
	\item The $a_{\mu}^{\rm SUSY}$ contribution is expected to be as large as possible in this work. This is ensured by defining the $\chi^2_{a_\mu}$ term as follows.
	\begin{equation}
		\chi^2_{a_\mu} =
		\begin{cases}
			1000\left(\frac{a_\mu^{\rm SUSY} - 2.68 \times 10^{-9}}{0.8\times10^{-9}}\right)^2, 	&a_\mu^{\rm SUSY} < 2.68\times10^{-9}\\
			0, &a_{\mu}^{\rm SUSY} \geq 2.68\times10^{-9}
		\end{cases}
	\end{equation}
	\item The $\chi^2_{\rm vev}$ term is introduced to ensure that sneutrino fields do not acquire non-zero vevs, according to the above-discussed assumption. Therefore, $\chi^2_{\rm vev} = 0$ if the electroweak vacuum of the parameter point is stable. Otherwise, $\chi^2_{\rm vev} = 10000$ if the electroweak vacuum is unstable or sneutrino fields attain non-zero vevs.
	\item The $\chi^2_{\rm veto}$ term is introduced to ensure that the LSP is a $\tau$-type sneutrino with $m_{\tilde{\nu}_1} > 300~{\rm GeV}$, according to the above-discussed assumption. Therefore, $\chi^2_{\rm veto} = 0$ if the parameter point satisfies this assumption; otherwise, $\chi^2_{\rm veto} = 10000$.
\end{itemize}
\par The GA method provides no unique solution to the minimization of $\chi^2$. Therefore, we selected the parameter space of one of the best solutions as a benchmark point for assessing the potential of the ISS-NMSSM to contribute a sufficiently large value of $a_\mu^{\rm SUSY}$ to account for $\Delta a_{\mu}$. Representative parameters and observables of the benchmark point are given in Table~\ref{tab:bp}.
\begin{table}[ht]
\centering
\begin{tabular}{cc|cc|cc}
\hline
 $\lambda$					&0.0173  				&$\kappa$							&0.0551 	&$\mu$							&534.6~GeV				\\
 $\tan{\beta}$				&59.90 					&$\lambda_{N_\mu}$					&0.1171		&$A_{\lambda_{N_\mu}}$			&2496.2~GeV				\\
 $y_\nu$  					&0.4804					&$A_{y_\nu}$						&111.8~GeV	&$a_{\mu}^{\rm SUSY}$  			&$2.318\times10^{-9}$  	\\
 $m_h$						&124.2~GeV  			&$m_{\tilde{\mu}_1}$				&720.2~GeV 	&$m_{\tilde{\mu}_2}$			&1446.4~GeV  			\\
 $m_{\widetilde{\chi}_1^0}$	&552.4~GeV  			&$m_{\widetilde{\chi}_2^0}$			&554.4~GeV  &$m_{\widetilde{\chi}_3^0}$		&2984.2~GeV 			\\
 $m_{\tilde{\nu}^R_1}$  	&418.2~GeV  			&$m_{\widetilde{\chi}_1^\pm}$		&553.7~GeV 	&$m_{\widetilde{\chi}_2^\pm}$	&2984.4~GeV  			\\
 $m_{\tilde{\nu}^I_1}$		&418.2~GeV 				&$m_{\tilde{\nu}^{I}_2}$  			&448.6~GeV	&$m_{\tilde{\nu}^{I}_3}$  		&931.0~GeV   			\\
 $Z^I_{1\sigma_{L}^\tau}$  	&$9.318\times10^{-3}$  	&$Z^I_{2\sigma_{L}^\mu}$  			&0.8949		&$Z^I_{3\sigma_{L}^\mu}$  		&0.4461					\\
 $Z^I_{1\sigma_{R}^\tau}$  	&0.7067  				&$Z^I_{2\sigma_{R}^\mu}$  			&0.3130 	&$Z^I_{3\sigma_{R}^\mu}$  		&0.6371  				\\
 $Z^I_{1\sigma_{x}^\tau}$  	&0.7074					&$Z^I_{2\sigma_{x}^\mu}$  			&0.3183  	&$Z^I_{3\sigma_{x}^\mu}$  		&0.6284  				\\
 $\Omega h^2$  				&0.1186 				&$\sigma^{\rm SI}_{\tilde{\nu}-p}$  &$4.713\times 10^{-48}~{\rm cm}^2$			&$\langle \sigma v \rangle_0$  	&$1.406\times10^{-29}~{\rm cm^3s^{-1}}$ 						\\
\hline
\end{tabular}
\caption{\label{tab:bp}Representative input parameters and observables of the benchmark point for assessing the potential of the ISS-NMSSM to contribute a sufficiently large value of $a_\mu^{\rm SUSY}$ to account for $\Delta a_{\mu}$.}
\end{table}
\par The results of the benchmark point in Table~\ref{tab:bp} indicate that the ISS-NMSSM can obtain a sufficiently large value of $a_{\mu}^{\rm SUSY}$ without contradicting the results of collider and DM direct detection experiments.
The results in Table~\ref{tab:bp} also indicate that the light neutralinos and charginos are Higgsino dominated and smuons are more massive than the $\mu$-type sneutrinos. Therefore, the charged HS loop provides a large contribution to $\Delta a_{\mu}$ in the form of $a_{\mu}^{\rm SUSY}$. Applying the parameters of the benchmark point to Eq.~(\ref{eq:simau}) indicates that a $\tan{\beta}$ enhancement effect is also needed in the HS contribution to $a_{\mu}^{\rm SUSY}$ because the muon Yukawa coupling $y_{\mu} = \frac{ m_\mu g_2}{\sqrt{2}m_W \cos{\beta}} \approx \frac{m_{\mu} g_2}{ \sqrt{2}m_W} \tan{\beta}$.
\par From the perspective of collider search, we note that a decreased $\mu$ is also required to ensure a sufficiently large value of $a_{\mu}^{\rm SUSY}$. As discussed above, the only visible channel representing the decay modes of Higgsinos is chargino pair production. However, the cross section of the pure Higgsino component of chargino pair production is less than the cross section of a pure Wino by a factor of about 3.5 \cite{Fuks:2013vua, LHC-SUSY-Xsect}. Here, a recent report from ATLAS cited a $95\%$ confidence level sensitivity to $m_{\widetilde{\chi}_1^{\pm}} = 1000~{\rm GeV}$ for a pure Wino using $139~{\rm fb^{-1}}$ data obtained through the $pp\to \widetilde{\chi}_1^{\pm}\widetilde{\chi}_1^{\mp} \to \tilde{\ell}\tilde{\ell}/ \tilde{\nu}\tilde{\nu} \to 2\ell + E_{\rm T}^{\rm miss}$ channel \cite{ATLAS-CONF-2019-008}. Assuming that the cross section of $\widetilde{\chi}_1^{\pm}$ pair production is unsuppressed and the acceptance rate and efficiency are unchanged for the $\tilde{\nu}$ LSP, the parameter space of the benchmark point remains outside of the exclusion range in the $m_{\widetilde{\chi}_1^{\pm}} - m_{\rm LSP}$ plane.

\par With respect to DM phenomenology, various annihilation mechanisms exist for sneutrino DM that predict the correct DM relic density. Of particular interest here is a co-annihilation mechanism with a Higgsino, which indicates that $\mu \simeq m_{\tilde{\nu}_1}$. As a consequence, leptons detected from the $2\ell + E_{\rm T}^{\rm miss}$ signal are too soft to be separated from background events, and cannot be detected by the LHC. Therefore, the constraint $m_{\tilde{\nu}_1} > 300~{\rm GeV}$ can be neglected in this compressed mass spectrum. This means that the mass of the $\mu$-type sneutrino can be as little as $100-200~{\rm GeV}$, which further increases the value of $a_{\mu}^{\rm SUSY}$. A detailed discussion of DM phenomenology in the ISS-NMSSM was presented in our previous work \cite{Cao:2017cjf}.

\subsection{Parameter features of muon \texorpdfstring{$g-2$}{} }
\begin{figure}[ht]

	\makebox[\textwidth][c]{	
	\subfigure[``Normal'' group]{\includegraphics[width=0.45\paperwidth]{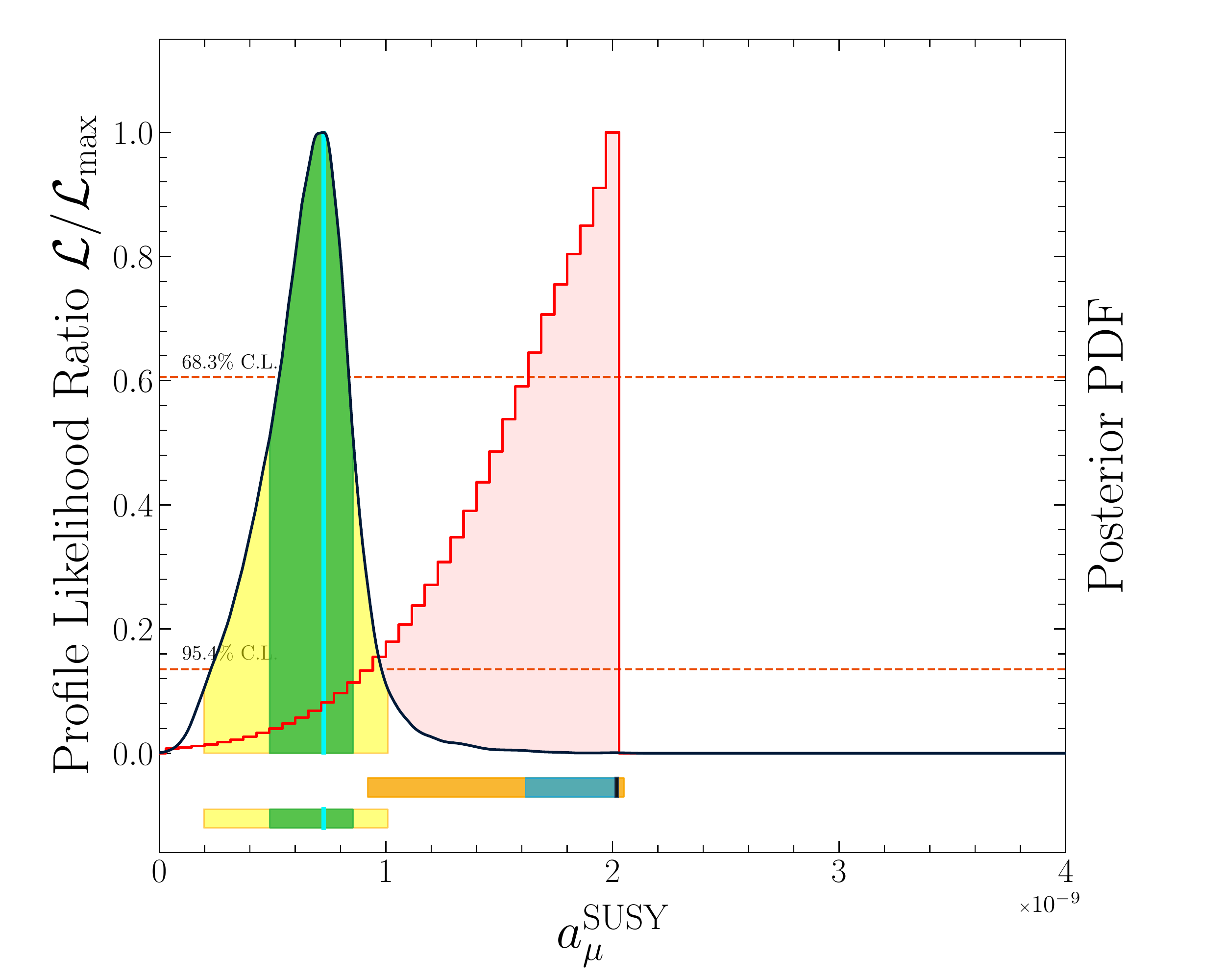}}\hspace{-0.8cm}
	\subfigure[``Control'' group]{\includegraphics[width=0.45\paperwidth]{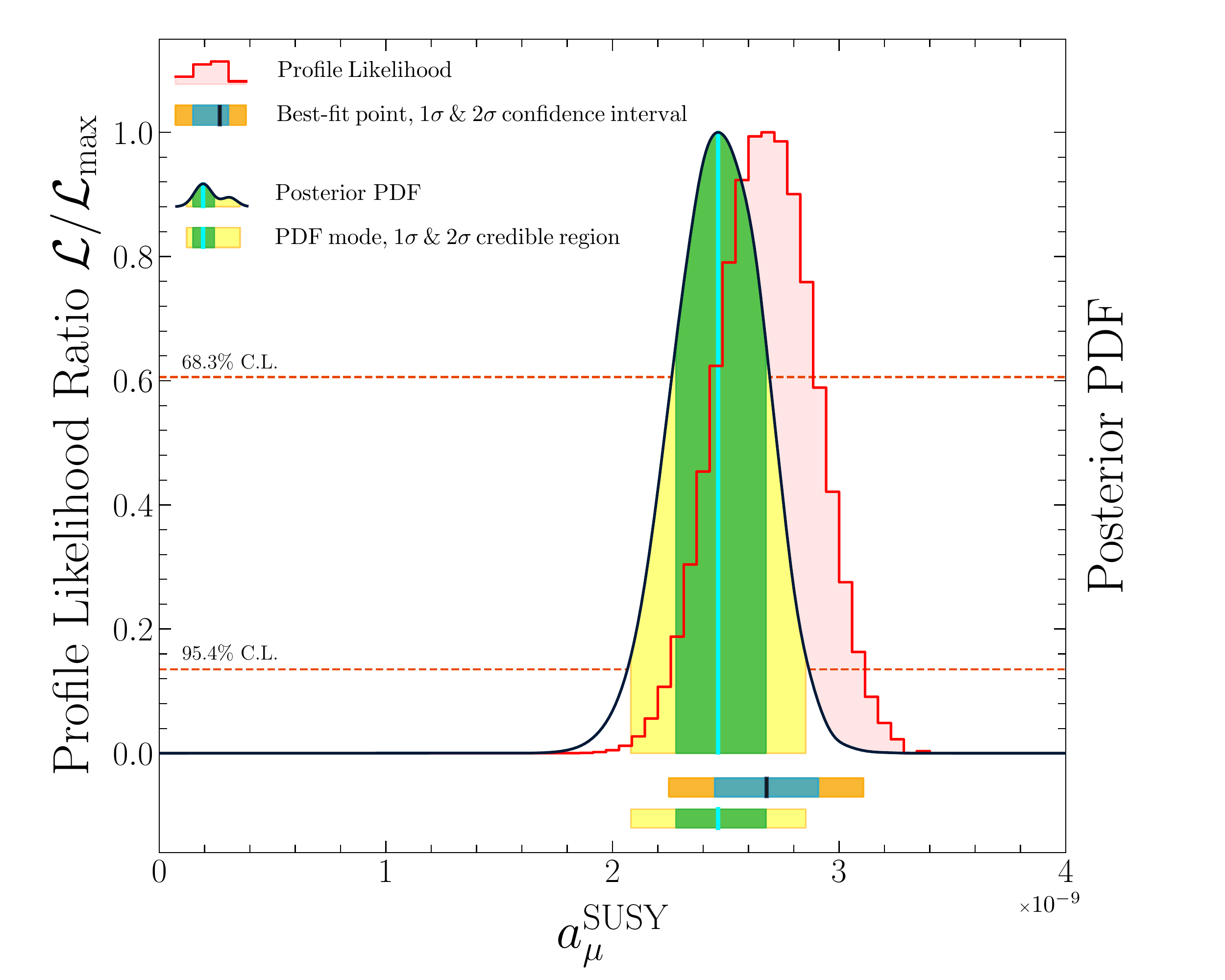}}
	}\\ \vspace{-0.3cm}
	
\caption{\label{fig:1damu} One dimensional profile likelihood and posterior PDF distributions as a function of $a_{\mu}^{\rm SUSY}$. The left panel is obtained by the result of the likelihood function in Eq.~(\ref{eq:liki}), while the right panel is for the likelihood function with $\delta a_\mu = 0.2 \times 10^{-9}$. Regions shaded with the blue (orange) color bar are the $1\sigma$ ($2\sigma$) confidence interval, in which the best-point is marked by the black vertical line. And those with the  green (yellow) color bar denote the $1\sigma$ ($2\sigma$) credible region.}
\end{figure}

\begin{figure}[t]
\vspace{-1cm}
	
	\makebox[\textwidth][c]{	
	\subfigure{\includegraphics[width=0.3\paperwidth]{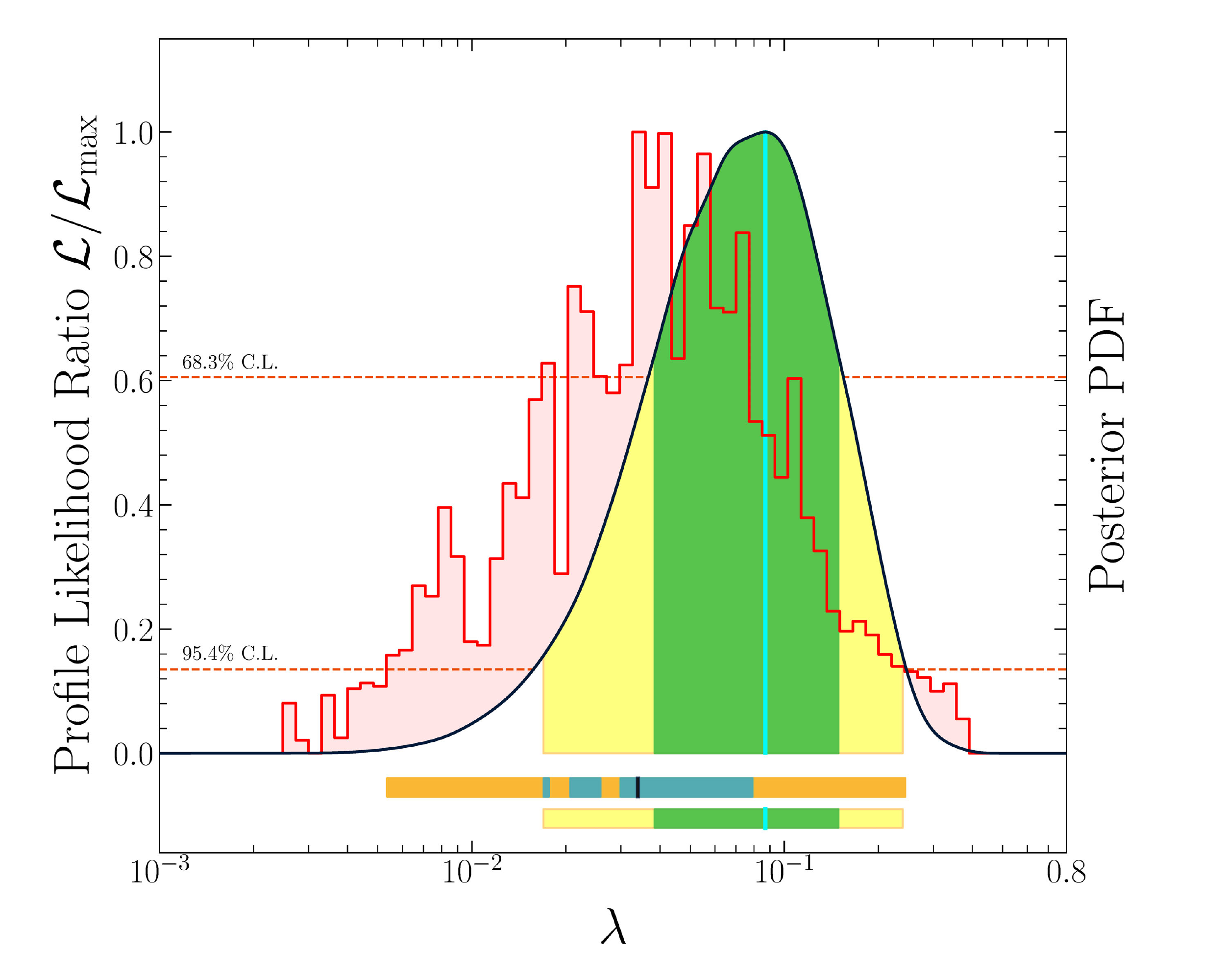}}\hspace{-0.4cm}
	\subfigure{\includegraphics[width=0.3\paperwidth]{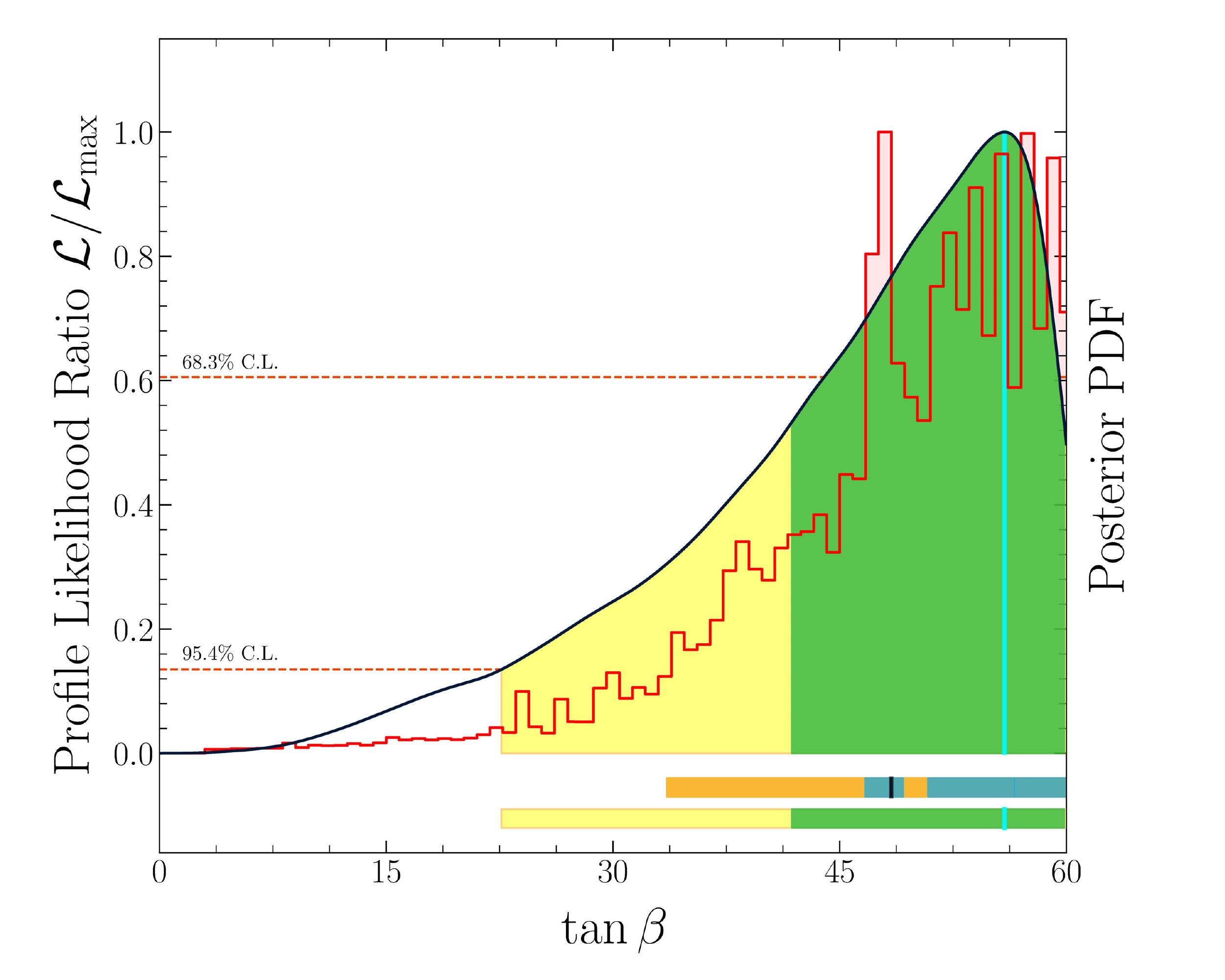}}\hspace{-0.4cm}
	\subfigure{\includegraphics[width=0.3\paperwidth]{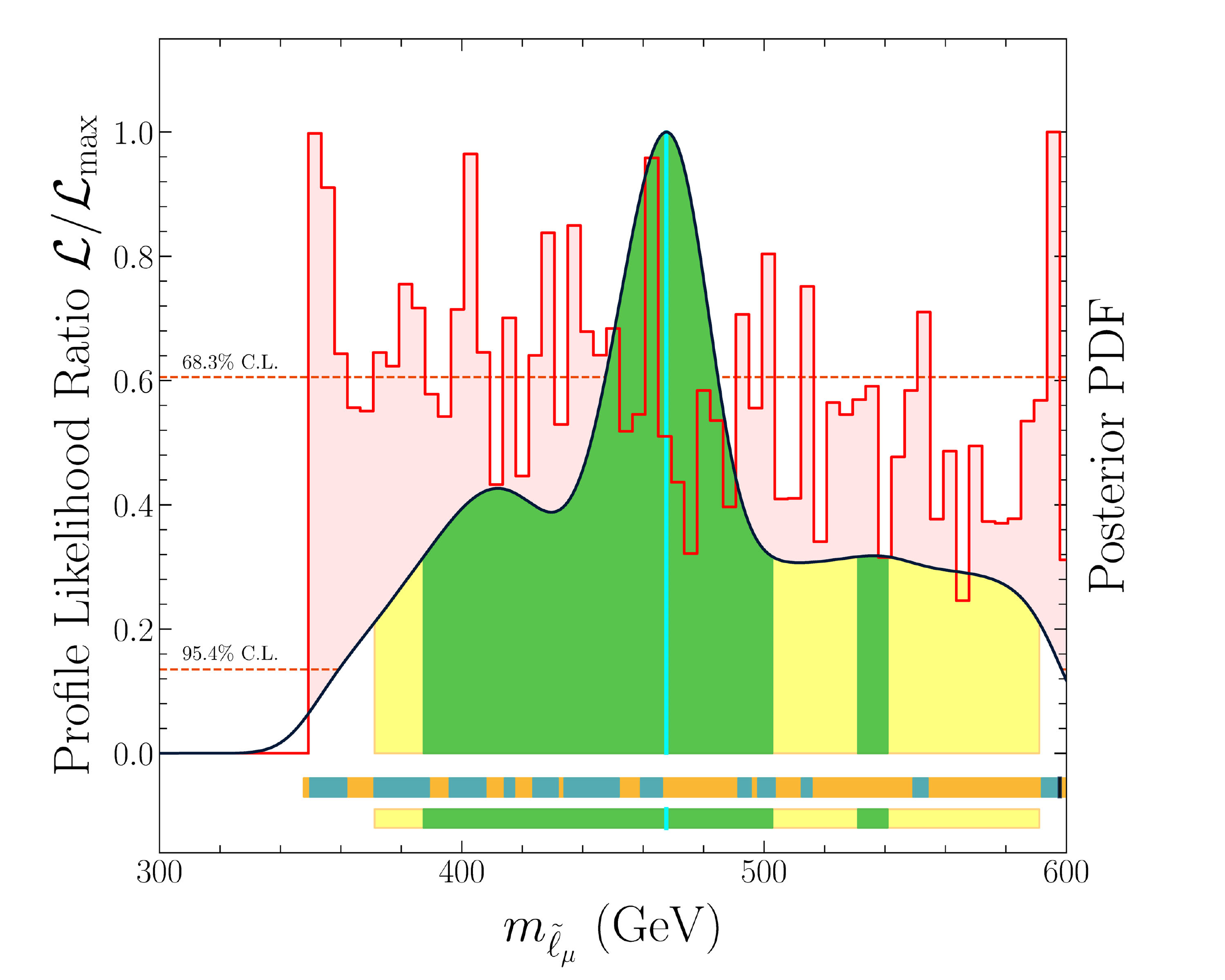}}
	}\\ \vspace{-0.6cm}	
	
	\makebox[\textwidth][c]{	
	\subfigure{\includegraphics[width=0.3\paperwidth]{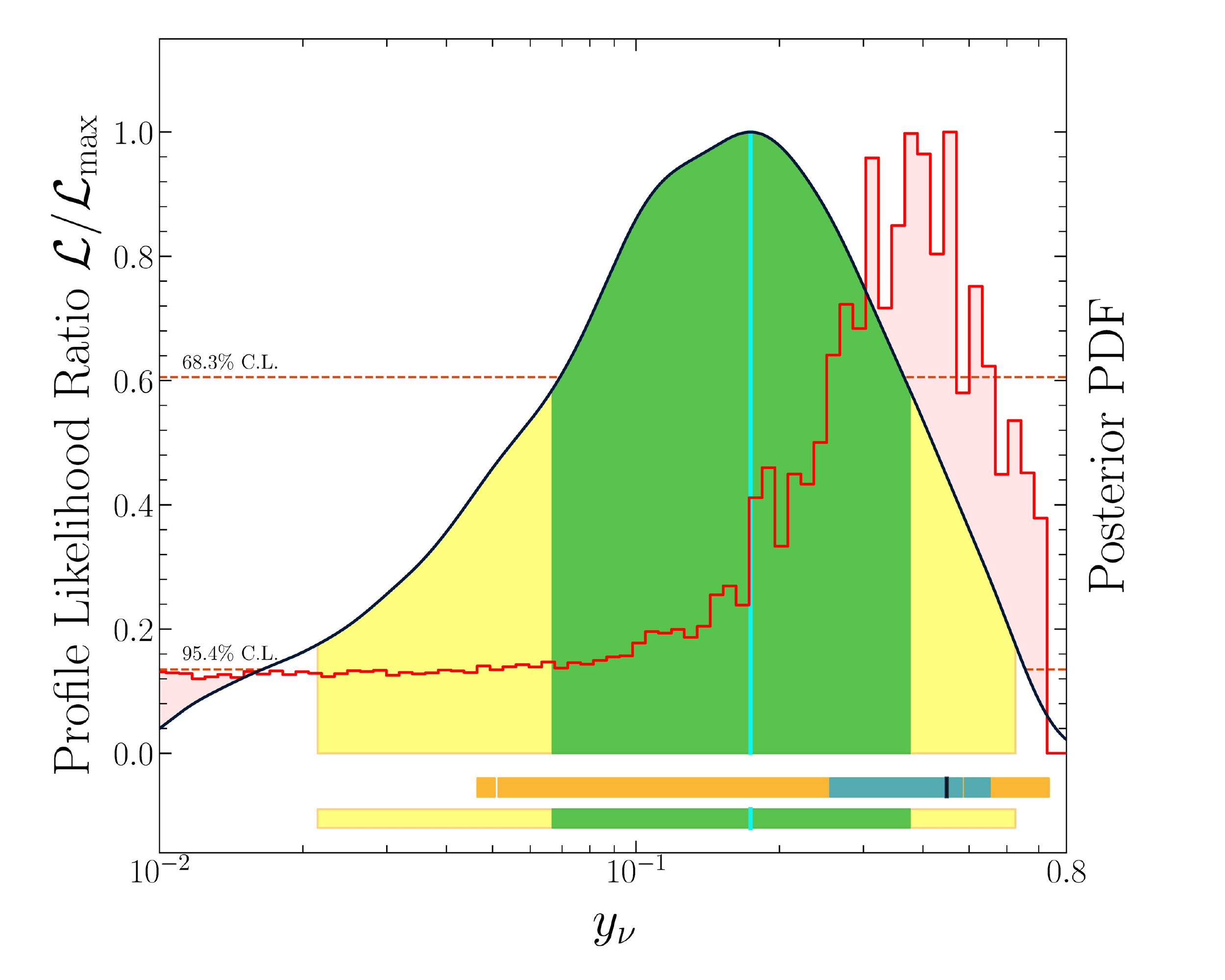}}\hspace{-0.4cm}
	\subfigure{\includegraphics[width=0.3\paperwidth]{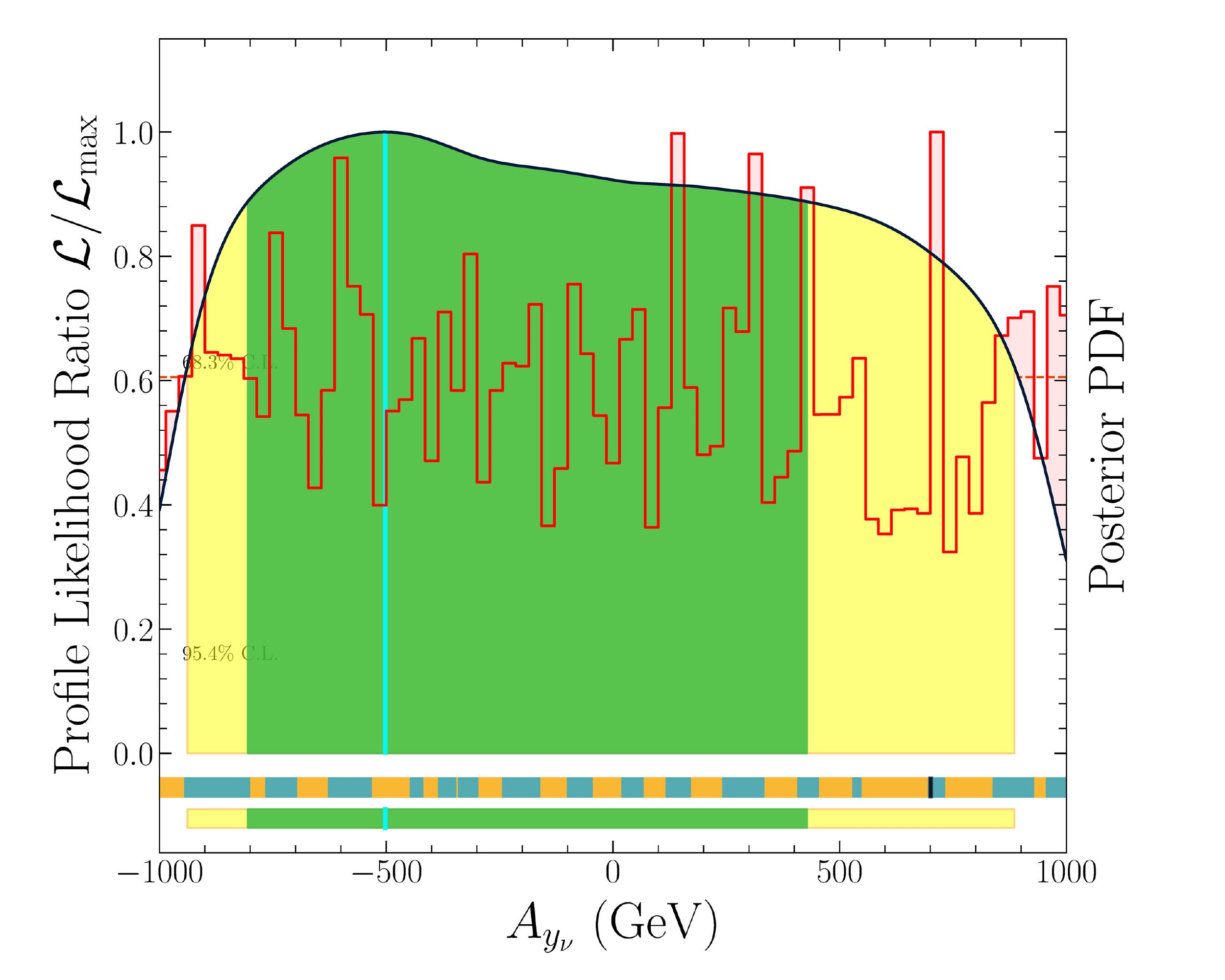}}\hspace{-0.4cm}
	\subfigure{\includegraphics[width=0.3\paperwidth]{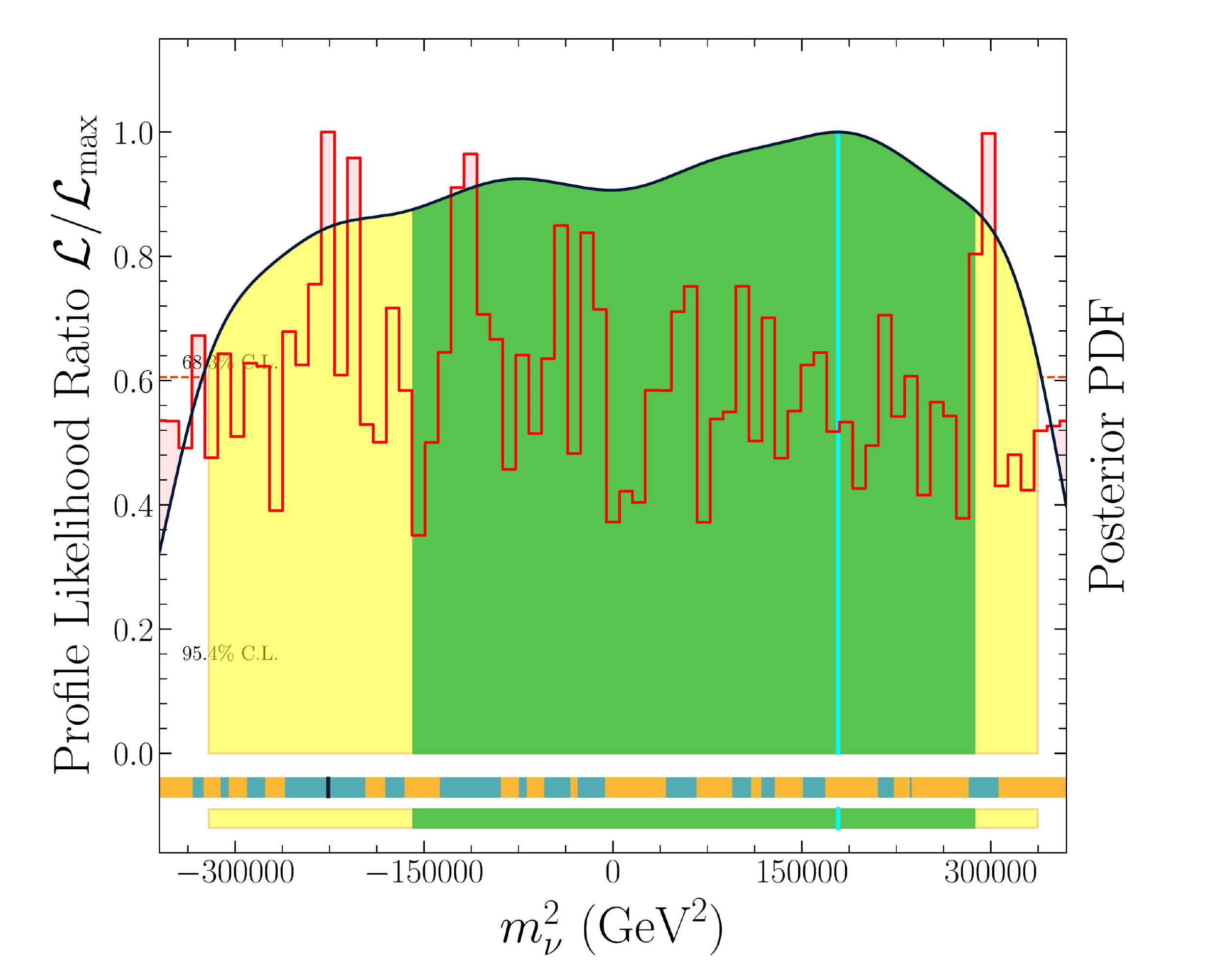}}
	}\\ \vspace{-0.6cm}

	\makebox[\textwidth][c]{	
	\subfigure{\includegraphics[width=0.3\paperwidth]{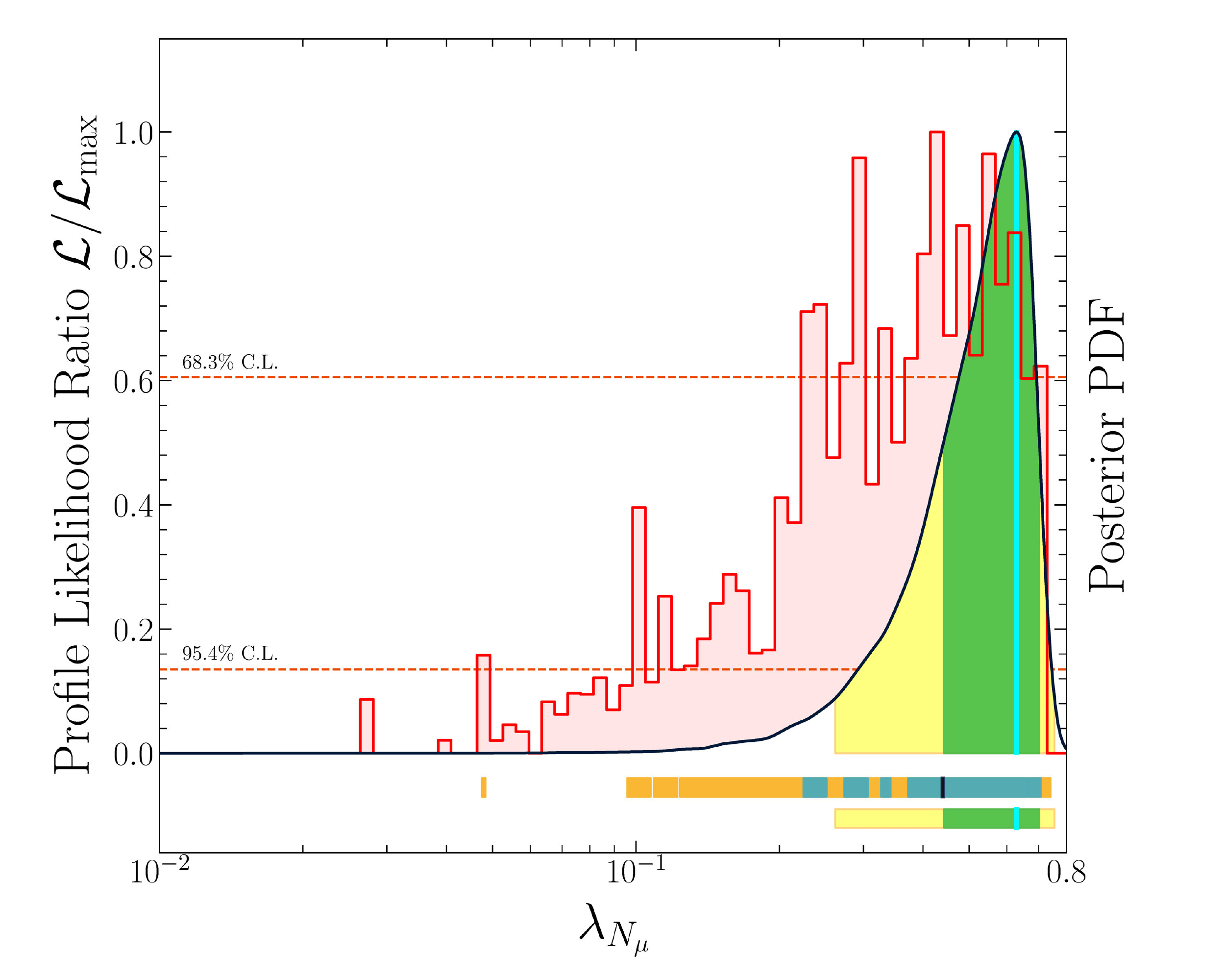}}\hspace{-0.4cm}
	\subfigure{\includegraphics[width=0.3\paperwidth]{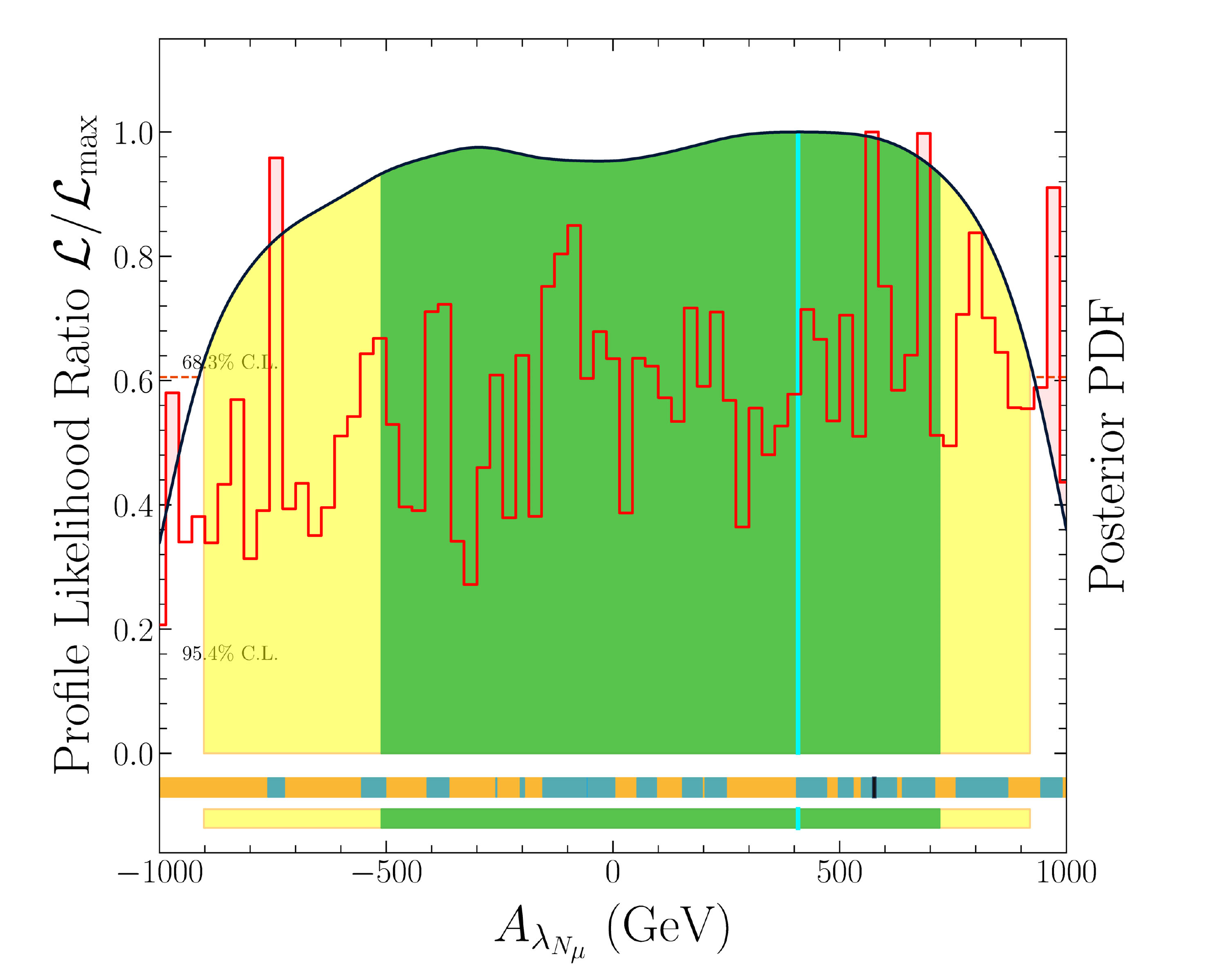}}\hspace{-0.4cm}
	\subfigure{\includegraphics[width=0.3\paperwidth]{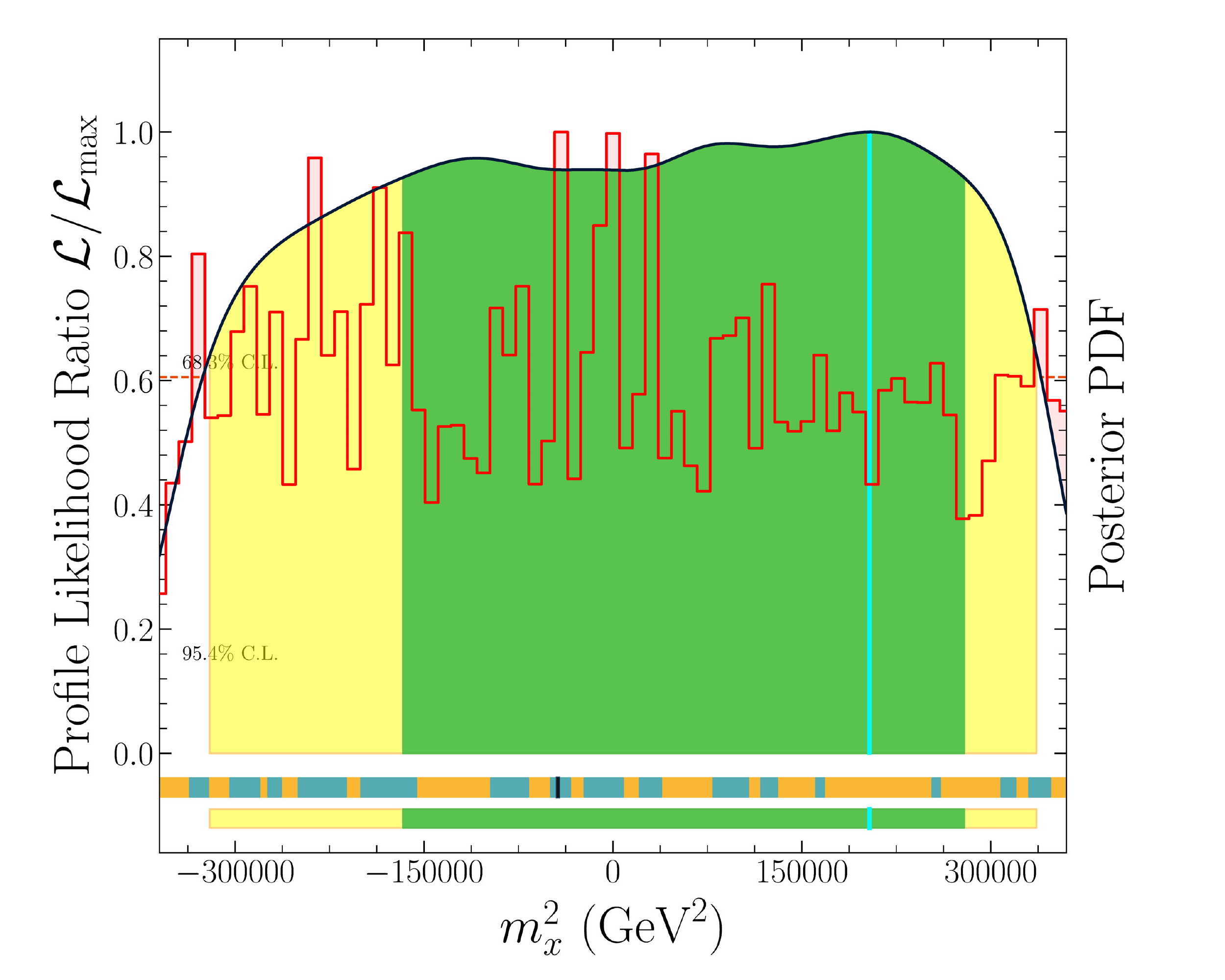}}
	}\\ \vspace{-0.8cm}

\caption{\label{fig:multipara} Similar to Fig.~\ref{fig:1damu}, but for the input parameters of ``Normal'' group.}
\end{figure}

The GA method cannot provide global information of $a_{\mu}^{\rm SUSY}$ in the parameter space. However, getting a correct statistical distribution without losing generality in such a high dimensional space in Eq.~(\ref{eq:para}) is a huge challenge for any scan algorithm. We note that, if the $\tau$-type sneutrino is approximately degenerate with Higgsino in mass, all DM measurements are easily satisfied by solely adjusting the parameters in $\tau$-type sneutrino sector. This motivates us to neglect the DM constraints by fixing the third generation slepton parameters and $\mu = 350~{\rm GeV}$ in studying $a_{\mu}^{\rm SUSY}$. Since the DM is massive ($m_{\tilde{\nu}_1} \simeq 350~{\rm GeV}$), the constraints from the sparticle searches at LHC are satisfied. In the following, we use the \texttt{MultiNest} sampling technique to scan the following parameter space:
	\begin{equation}
	\begin{split}
	&0.001 < \lambda < 0.7,\quad |\kappa| < 0.7,\quad 1<\tan{\beta}< 60,\quad \mu=350~{\rm GeV}, \\
	&|A_\kappa| < 1~{\rm TeV}, \quad |A_t|<5~{\rm TeV},\quad A_t=A_b,\quad A_\lambda = 2~{\rm TeV},\\
	&0.01 < y_\nu < 0.7,\quad |A_{y_{\nu}}| < 1~{\rm TeV}, \quad 0.01 < \lambda_{N_\mu} < 0.7, \quad \left|A_{\lambda_{N_\mu}} \right| < 1~{\rm TeV}, \\
	& 350~{\rm GeV} < m_{\tilde{\ell}_\mu} < 600~{\rm GeV}, \quad \left|m_{\nu}^2\right| < (800~{\rm GeV})^2, \quad \left| m_{x}^2 \right| < (800~{\rm GeV})^2.
	\end{split}
	\end{equation}	
The prior probability distribution function (PDF) of these inputs are setted as uniformly distributed and the $n_{\rm live}$\footnote{In the \texttt{MultiNest} algorithm, $n_{\rm live}$ represents the number of active or live points used to determine the iso-likelihood contour in each iteration~\cite{Feroz:2008xx, Feroz:2013hea}} parameter is setted at 10000. The likelihood function adopted in the scan is a standard Gaussian form of muon $(g-2)$:
	\begin{equation}\label{eq:liki}
	\mathcal{L} = \exp\left[ -\frac{1}{2} \left(\frac{a_{\mu}^{\rm SUSY} - 2.68\times10^{-9}}{0.8\times10^{-9}}\right)^2 \right].
	\end{equation}
	During the scan, we require the lightest $\mu$-type sneutrino mass is larger than $350~{\rm GeV}$, the neutrino unitary bound is satisfied and any sneutrino field is forbidden to develop a vev. Besides, only the samples consistent with the discovered SM-like Higgs boson data are retained, and the consistency is checked by code \texttt{HiggsSignals}.	 

\begin{table}[]
\centering
\resizebox{0.95\textwidth}{!}{
\begin{tabular}{c|c|c}
\hline
\hline
\multirow{3}{*}{Parameters or observables} & \multicolumn{2}{c}{$1\sigma$ $2\sigma$ credible regions} \\ \cline{2-3} 
                                          & Normal               & Control              \\
                                          & $\mathcal{L} = \exp\left[ -\frac{1}{2} \left(\frac{a_{\mu}^{\rm SUSY} - 2.68\times10^{-9}}{0.8\times10^{-9}}\right)^2 \right]$                 & $\mathcal{L} = \exp\left[ -\frac{1}{2} \left(\frac{a_{\mu}^{\rm SUSY} - 2.68\times10^{-9}}{0.2\times10^{-9}}\right)^2 \right]$                  \\ \hline
  $a_{\mu}^{\rm SUSY}/10^{-9}$   &   $[0.48, 0.86]~~~~~[0.19, 1.01]$                   & $[2.28, 2.67]~~~~~[2.08, 2.85]$ \\
  $\lambda$                      &   $[0.038, 0.15]~~~~~[0.017, 0.24]$                   &  $[0.027, 0.055]~~~~~[0.017, 0.063]$\\
  $\tan{\beta}$                  &   $[42, 60]~~~~~[23, 60]$                   
&  $[55, 60]~~~~~[51, 60]$\\
  $y_\nu$                        &  $[0.066, 0.37]~~~~~[0.022, 0.63]$ 
&  $[0.41, 0.55]~~~~~[0.36, 0.64]$ \\
  $A_{y_{\nu}}/{\rm GeV}$                  &  $[-824, 407]~~~~~[-963, 864]$ 
&  $[-423, 753]~~~~~[-855, 928]$ \\
  $\lambda_{N_{\mu}}$            &  $[0.43, 0.69]~~~~~[0.26, 0.70]$ 
&  $[0.37, 0.66]~~~~~[0.25, 0.72]$ \\                                                                                    
  $A_{\lambda_{N_{\mu}}}/{\rm GeV}$        &  $[-516, 715]~~~~~[-911, 916]$   
& $[256, 943 ]~~~~~[-378, 1000]$\\
  $m_{\tilde{\ell}_\mu}/{\rm GeV}$         &  $[388, 502]~~~~~[371, 590]$  
& $[358, 473]~~~~~[350, 558]$\\
  $m_{\nu}^2/(10^3~{\rm GeV}^2)$            &  $[-162, 284]~~~~~[-323, 335]$ & $[-277, 188]~~~~~[-335, 317]$\\
  $m_{x}^2/(10^3~{\rm GeV}^2)$              &  $[-167, 278]~~~~~[-320, 335]$ & $[-221, 246]~~~~~[-325, 327]$ \\
\hline\hline

\end{tabular}}
\caption{\label{tab:1sigma}One-dimensional credible regions of $a_{\mu}^{\rm SUSY}$ and input parameters. The intervals in the first and second brackets correspond to the $1\sigma$ and $2\sigma$ credible regions respectively. }
\end{table}

\par This scan is marked as ``Normal'' group, and its the one-dimensional profile likelihood (PL) of the $\mathcal{L}$ in Eq.~(\ref{eq:liki}) and one-dimensional marginal posterior PDF for $a_{\mu}^{\rm SUSY}$ and the related input parameters are plotted in Fig.~{\ref{fig:1damu}~(a) and Fig.~{\ref{fig:multipara}. The one-dimensional PL of an interested parameter or an observable $\theta$ on position $\theta=\theta_0$ is defined as the maximum value of $\mathcal{L}$:
\begin{equation}
	\mathcal{L}(\theta_0) = \max\left(  \mathcal{L}|_{\theta=\theta_0} \right),
\end{equation} 
where the maximization is through varying the other input parameters. PL can be viewed as an local predictive capability indicator of the theory. Consequently, the best point in the sample corresponding to the peak position of PL $\mathcal{L}_{\rm max}$. Complementarity, the one-dimensional marginal posterior PDF is a global statical quantity. 

Fig.~{\ref{fig:1damu}~(a) indicates that the magnitude of HS contribution concentrates around $7\times 10^{-10}$ for the ``Normal'' group, and approximately $3\%$ of the samples obtained results with $a_{\mu}^{\rm SUSY} > 10^{-9}$ (see the black curve). Fig.~\ref{fig:multipara} shows that a small $\lambda$, a large $\tan{\beta}$ and a large $y_\nu$ are favored when predicting a relatively large $a_{\mu}^{\rm SUSY}$; the plots of $y_\nu$ and $\lambda_{N_\mu}$ confirm that the unitary condition in Eq.~(\ref{eq:uc}) usually limits a large $y_\nu$. By contrast, the PL has no particular preference on the parameters $A_{y_\nu}$, $A_{\lambda_{N_{\mu}}}$ $m_{\nu}^2$ and $m_{x}^2$ (see the red step line).

Whether the HS contribution alone can explain $\Delta a_{\mu}$ is particularly interested, so we also carried out a comparative ``Control'' group scan, which is same as the ``Normal'' scan except for the replacement of the uncertainty $0.8\times 10^{-9}$ in Eq.~(\ref{eq:liki}) by $0.2 \times 10^{-9}$.  The distributions of $a_{\mu}^{\rm SUSY}$ is presented in Fig.~{\ref{fig:1damu}~(b). This panel shows that there is a certain range of parameter space in ISS-NMSSM where the HS contribution alone can explain $\Delta a_{\mu}$. In practice, the ``Control'' scan consumes more computing resource than the ``Normal'' group. The underlying reason is that, in order to predict a larger $a_{\mu}^{\rm SUSY}$, a much more fine-tuned parameter configuration is necessary, so the samples of the ``Control'' group are more harder to obtain.

\par For completeness, the $1\sigma$ and $2\sigma$ credible regions of both ``Normal''  and ``Control'' scans are summarized in Table~\ref{tab:1sigma}. It is evident that, the regions are quite different due to the different choices of the uncertainty.

\subsection{Electron \texorpdfstring{$g-2$}{} in ISS-NMSSM}
In addition to $\Delta a_{\mu}$, we also note that about a $2.5\sigma$ discrepancy has been reported between the experimental observations and the SM prediction for the anomalous electron magnetic moment $\Delta a_e = -0.88(36)\times10^{-12}$ \cite{Hanneke:2008tm, Hanneke:2010au}. This leads to the question as to whether the ISS-NMSSM can account for the observed $\Delta a_{\mu}$ and $\Delta a_e$ simultaneously. 
Concerning this question, the SUSY contribution to lepton anomalous magnetic moment $a_{\ell}^{\rm SUSY}~(\ell = e$ or $\mu)$ can be factorized into the lepton mass square times a SUSY factor $R_\ell$ if there is no flavor mixing in slepton sector. In the current situation, $R_\ell$ for electron and muon are 
	\begin{equation}
	\begin{split}
		R_{e} = \frac{\Delta a_e}{m_e^2} &= \frac{-0.88\times 10^{-12}}{(0.511\times 10^{-3}~{\rm GeV})^2} = -3.370\times10^{-6}~{\rm GeV}^{-2}, \\
		R_{\mu} = \frac{\Delta a_{\mu}}{m_\mu^2} &= \frac{268\times 10^{-11}}{(1.057\times 10^{-1}~{\rm GeV})^2} = 2.399\times 10^{-7}~{\rm GeV}^{-2}.
	\end{split}
	\end{equation}
	This difference of $-14$ between $R_e$ and $R_\mu$ indicates the two anomalies are hard to explain by a common physical origin. A recent unified explanation of the discrepancies was studies in the MSSM~\cite{Badziak:2019gaf}, and the critical points for the solution are as follows:
\begin{itemize}
\item The Bino-selectron loop is responsible for $\Delta a_e$, which needs moderately small Bino and selectron masses and $\mu M_1 < 0$.
\item The Wino-sneutrino loop accounts for $\Delta a_\mu$, which essentially requires $\mu M_2  > 0$.
\end{itemize}

From the discussion of $\Delta a_\mu$ in this work, one can infer the following conclusions for the ISS-NMSSM:
\begin{itemize}
\item As indicated by Eq.~(\ref{eq:damufull})-(\ref{eq:simau}), a negative $\Delta a_e$ is obtainable from the HS contribution if the rotation matrix for the $e$-type sneutrino
fields has the property ${\rm sgn}({Z_{n2}^{*} Z_{n1}}) = -1$. This condition can be satisfied by flipping the sign of $A_{Y_{\nu_e}}$ in the chiral sneutrino mixing term.
\item As shown in Eq.~(\ref{eq:uc}), the neutrino unitary constrain on $Y_{\nu_e}$ is significantly weaker than that on $Y_{\nu_\mu}$. So the HS contribution can predict a relatively larger $|R_e|$.
\item Although it is unlikely for the HS contribution alone to reconcile both discrepancies when $Y_{\nu_e}$ and $|A_{Y_{\nu_e}}|$ are not tremendously large, the tension between theory and experiment can be relaxed significantly, such as the special requirements for the signs of $M_1$, $M_2$ and $\mu$. So comparing with MSSM framework~\cite{Badziak:2019gaf}, explaining the discrepancies simultaneously is more easier by the other contributions of the ISS-NMSSM.
\end{itemize}

\section{Summary}\label{sec:sum}
In this work, we performed a detailed phenomenological study of the anomalous muon magnetic moment $a_{\mu}$ in the ISS-NMSSM. The results demonstrated that the newly introduced Yukawa coupling $Y_\nu$ in the ISS-NMSSM significantly increased the value of $a_{\mu}$, relative to that obtained with the standard NMSSM, via a mixing between left-handed and right-handed sneutrinos in the chargino-sneutrino loop diagram. Moreover, the right-handed sneutrino serves as a good DM candidate with an undetectable DM-nucleus scattering rate, and the constraints arising from the LHC, $B$-physics observations, and Higgs global fitting can also be naturally satisfied.

Accordingly, if the statistically significant deviation of $a_\mu$ between experimental observation and SM prediction confirmed by the upcoming Fermilab E989 experiment and theoretical studies, ISS-NMSSM may be a better electroweak SUSY framework.

\acknowledgments
Pengxuan Zhu wishes to thank Dr. Yusi Pan for the helpful discussion. This work was supported by the National Natural Science Foundation
of China (NNSFC) under Grant No. 11575053.

\providecommand{\href}[2]{#2}\begingroup\raggedright\endgroup


\begin{thebibliography}{100}

\bibitem{Bennett:2006fi}
{\scshape Muon g-2} collaboration, G.~W. Bennett et~al., \emph{{Final Report of
  the Muon E821 Anomalous Magnetic Moment Measurement at BNL}},
  \href{https://doi.org/10.1103/PhysRevD.73.072003}{\emph{Phys. Rev.}
  {\bfseries D73} (2006) 072003}
  [\href{https://arxiv.org/abs/hep-ex/0602035}{{\ttfamily hep-ex/0602035}}].

\bibitem{Blum:2013xva}
T.~Blum, A.~Denig, I.~Logashenko, E.~de~Rafael, B.~L. Roberts, T.~Teubner
  et~al., \emph{{The Muon (g-2) Theory Value: Present and Future}},
  \href{https://arxiv.org/abs/1311.2198}{{\ttfamily 1311.2198}}.

\bibitem{Keshavarzi:2018mgv}
A.~Keshavarzi, D.~Nomura and T.~Teubner, \emph{{Muon $g-2$ and $\alpha(M_Z^2)$:
  a new data-based analysis}},
  \href{https://doi.org/10.1103/PhysRevD.97.114025}{\emph{Phys. Rev.}
  {\bfseries D97} (2018) 114025}
  [\href{https://arxiv.org/abs/1802.02995}{{\ttfamily 1802.02995}}].

\bibitem{Aoyama:2012wk}
T.~Aoyama, M.~Hayakawa, T.~Kinoshita and M.~Nio, \emph{{Complete Tenth-Order
  QED Contribution to the Muon g-2}},
  \href{https://doi.org/10.1103/PhysRevLett.109.111808}{\emph{Phys. Rev. Lett.}
  {\bfseries 109} (2012) 111808}
  [\href{https://arxiv.org/abs/1205.5370}{{\ttfamily 1205.5370}}].

\bibitem{Aoyama:2017uqe}
T.~Aoyama, T.~Kinoshita and M.~Nio, \emph{{Revised and Improved Value of the
  QED Tenth-Order Electron Anomalous Magnetic Moment}},
  \href{https://doi.org/10.1103/PhysRevD.97.036001}{\emph{Phys. Rev.}
  {\bfseries D97} (2018) 036001}
  [\href{https://arxiv.org/abs/1712.06060}{{\ttfamily 1712.06060}}].

\bibitem{Gnendiger:2013pva}
C.~Gnendiger, D.~Stöckinger and H.~Stöckinger-Kim, \emph{{The electroweak
  contributions to $(g-2)_\mu$ after the Higgs boson mass measurement}},
  \href{https://doi.org/10.1103/PhysRevD.88.053005}{\emph{Phys. Rev.}
  {\bfseries D88} (2013) 053005}
  [\href{https://arxiv.org/abs/1306.5546}{{\ttfamily 1306.5546}}].

\bibitem{Nyffeler:2016gnb}
A.~Nyffeler, \emph{{Precision of a data-driven estimate of hadronic
  light-by-light scattering in the muon $g-2$: Pseudoscalar-pole
  contribution}}, \href{https://doi.org/10.1103/PhysRevD.94.053006}{\emph{Phys.
  Rev.} {\bfseries D94} (2016) 053006}
  [\href{https://arxiv.org/abs/1602.03398}{{\ttfamily 1602.03398}}].

\bibitem{Davier:2019can}
M.~Davier, A.~Hoecker, B.~Malaescu and Z.~Zhang, \emph{{A new evaluation of the
  hadronic vacuum polarisation contributions to the muon anomalous magnetic
  moment and to $\mathbf{\boldsymbol\alpha(m_Z^2)}$}},
  \href{https://doi.org/10.1140/epjc/s10052-020-7792-2}{\emph{Eur. Phys. J. C}
  {\bfseries 80} (2020) 241}
  [\href{https://arxiv.org/abs/1908.00921}{{\ttfamily 1908.00921}}].

\bibitem{Davier:2017zfy}
M.~Davier, A.~Hoecker, B.~Malaescu and Z.~Zhang, \emph{{Reevaluation of the
  hadronic vacuum polarisation contributions to the Standard Model predictions
  of the muon $g-2$ and ${\alpha (m_Z^2)}$ using newest hadronic cross-section
  data}}, \href{https://doi.org/10.1140/epjc/s10052-017-5161-6}{\emph{Eur.
  Phys. J.} {\bfseries C77} (2017) 827}
  [\href{https://arxiv.org/abs/1706.09436}{{\ttfamily 1706.09436}}].

\bibitem{Tanabashi:2018oca}
{\scshape Particle Data Group} collaboration, M.~Tanabashi et~al.,
  \emph{{Review of Particle Physics}},
  \href{https://doi.org/10.1103/PhysRevD.98.030001}{\emph{Phys. Rev.}
  {\bfseries D98} (2018) 030001}.

\bibitem{Davier:2010nc}
M.~Davier, A.~Hoecker, B.~Malaescu and Z.~Zhang, \emph{{Reevaluation of the
  Hadronic Contributions to the Muon g-2 and to alpha(MZ)}},
  \href{https://doi.org/10.1140/epjc/s10052-012-1874-8,
  10.1140/epjc/s10052-010-1515-z}{\emph{Eur. Phys. J.} {\bfseries C71} (2011)
  1515} [\href{https://arxiv.org/abs/1010.4180}{{\ttfamily 1010.4180}}].

\bibitem{Davier:2013sfa}
M.~Davier, A.~Höcker, B.~Malaescu, C.-Z. Yuan and Z.~Zhang, \emph{{Update of
  the ALEPH non-strange spectral functions from hadronic $\tau$ decays}},
  \href{https://doi.org/10.1140/epjc/s10052-014-2803-9}{\emph{Eur. Phys. J.}
  {\bfseries C74} (2014) 2803}
  [\href{https://arxiv.org/abs/1312.1501}{{\ttfamily 1312.1501}}].

\bibitem{Hagiwara:2011af}
K.~Hagiwara, R.~Liao, A.~D. Martin, D.~Nomura and T.~Teubner,
  \emph{{$(g-2)_\mu$ and $\alpha(M_Z^2)$ re-evaluated using new precise data}},
  \href{https://doi.org/10.1088/0954-3899/38/8/085003}{\emph{J. Phys.}
  {\bfseries G38} (2011) 085003}
  [\href{https://arxiv.org/abs/1105.3149}{{\ttfamily 1105.3149}}].

\bibitem{Jegerlehner:2009ry}
F.~Jegerlehner and A.~Nyffeler, \emph{{The Muon g-2}},
  \href{https://doi.org/10.1016/j.physrep.2009.04.003}{\emph{Phys. Rept.}
  {\bfseries 477} (2009) 1} [\href{https://arxiv.org/abs/0902.3360}{{\ttfamily
  0902.3360}}].

\bibitem{Beskidt:2012sk}
C.~Beskidt, W.~de~Boer, D.~I. Kazakov and F.~Ratnikov, \emph{{Constraints on
  Supersymmetry from LHC data on SUSY searches and Higgs bosons combined with
  cosmology and direct dark matter searches}},
  \href{https://doi.org/10.1140/epjc/s10052-012-2166-z}{\emph{Eur. Phys. J.}
  {\bfseries C72} (2012) 2166}
  [\href{https://arxiv.org/abs/1207.3185}{{\ttfamily 1207.3185}}].

\bibitem{Benayoun:2011mm}
M.~Benayoun, P.~David, L.~DelBuono and F.~Jegerlehner, \emph{{Upgraded Breaking
  Of The HLS Model: A Full Solution to the $\tau^-e^+e^-$ and $\phi$ Decay
  Issues And Its Consequences On g-2 VMD Estimates}},
  \href{https://doi.org/10.1140/epjc/s10052-011-1848-2}{\emph{Eur. Phys. J. C}
  {\bfseries 72} (2012) 1848}
  [\href{https://arxiv.org/abs/1106.1315}{{\ttfamily 1106.1315}}].

\bibitem{Benayoun:2012wc}
M.~Benayoun, P.~David, L.~DelBuono and F.~Jegerlehner, \emph{{An Update of the
  HLS Estimate of the Muon g-2}},
  \href{https://doi.org/10.1140/epjc/s10052-013-2453-3}{\emph{Eur. Phys. J. C}
  {\bfseries 73} (2013) 2453}
  [\href{https://arxiv.org/abs/1210.7184}{{\ttfamily 1210.7184}}].

\bibitem{Benayoun:2015gxa}
M.~Benayoun, P.~David, L.~DelBuono and F.~Jegerlehner, \emph{{Muon $g-2$
  estimates: can one trust effective Lagrangians and global fits?}},
  \href{https://doi.org/10.1140/epjc/s10052-015-3830-x}{\emph{Eur. Phys. J. C}
  {\bfseries 75} (2015) 613}
  [\href{https://arxiv.org/abs/1507.02943}{{\ttfamily 1507.02943}}].

\bibitem{Grange:2015fou}
{\scshape Muon g-2} collaboration, J.~Grange et~al., \emph{{Muon (g-2)
  Technical Design Report}},
  \href{https://arxiv.org/abs/1501.06858}{{\ttfamily 1501.06858}}.

\bibitem{Altmannshofer:2016brv}
W.~Altmannshofer, C.-Y. Chen, P.~S. Bhupal~Dev and A.~Soni, \emph{{Lepton
  flavor violating $Z^\prime$ explanation of the muon anomalous magnetic
  moment}}, \href{https://doi.org/10.1016/j.physletb.2016.09.046}{\emph{Phys.
  Lett.} {\bfseries B762} (2016) 389}
  [\href{https://arxiv.org/abs/1607.06832}{{\ttfamily 1607.06832}}].

\bibitem{Krnjaic:2019rsv}
G.~Krnjaic, G.~Marques-Tavares, D.~Redigolo and K.~Tobioka, \emph{{Probing
  Muonic Forces and Dark Matter at Kaon Factories}},
  \href{https://arxiv.org/abs/1902.07715}{{\ttfamily 1902.07715}}.

\bibitem{Darme:2018hqg}
L.~Darmé, K.~Kowalska, L.~Roszkowski and E.~M. Sessolo, \emph{{Flavor
  anomalies and dark matter in SUSY with an extra U(1)}},
  \href{https://doi.org/10.1007/JHEP10(2018)052}{\emph{JHEP} {\bfseries 10}
  (2018) 052} [\href{https://arxiv.org/abs/1806.06036}{{\ttfamily
  1806.06036}}].

\bibitem{Altmannshofer:2014pba}
W.~Altmannshofer, S.~Gori, M.~Pospelov and I.~Yavin, \emph{{Neutrino Trident
  Production: A Powerful Probe of New Physics with Neutrino Beams}},
  \href{https://doi.org/10.1103/PhysRevLett.113.091801}{\emph{Phys. Rev. Lett.}
  {\bfseries 113} (2014) 091801}
  [\href{https://arxiv.org/abs/1406.2332}{{\ttfamily 1406.2332}}].

\bibitem{Foldenauer:2018zrz}
P.~Foldenauer, \emph{{Light dark matter in a gauged $U(1)_{L_\mu-L_\tau}$
  model}}, \href{https://doi.org/10.1103/PhysRevD.99.035007}{\emph{Phys. Rev.}
  {\bfseries D99} (2019) 035007}
  [\href{https://arxiv.org/abs/1808.03647}{{\ttfamily 1808.03647}}].

\bibitem{Davoudiasl:2014kua}
H.~Davoudiasl, H.-S. Lee and W.~J. Marciano, \emph{{Muon $g-2$, rare kaon
  decays, and parity violation from dark bosons}},
  \href{https://doi.org/10.1103/PhysRevD.89.095006}{\emph{Phys. Rev.}
  {\bfseries D89} (2014) 095006}
  [\href{https://arxiv.org/abs/1402.3620}{{\ttfamily 1402.3620}}].

\bibitem{Mohlabeng:2019vrz}
G.~Mohlabeng, \emph{{Revisiting the dark photon explanation of the muon
  anomalous magnetic moment}},
  \href{https://doi.org/10.1103/PhysRevD.99.115001}{\emph{Phys. Rev.}
  {\bfseries D99} (2019) 115001}
  [\href{https://arxiv.org/abs/1902.05075}{{\ttfamily 1902.05075}}].

\bibitem{Fuyuto:2019vfe}
K.~Fuyuto, X.-G. He, G.~Li and M.~Ramsey-Musolf, \emph{{CP-violating Dark
  Photon Interaction}},  \href{https://arxiv.org/abs/1902.10340}{{\ttfamily
  1902.10340}}.

\bibitem{Bauer:2018onh}
M.~Bauer, P.~Foldenauer and J.~Jaeckel, \emph{{Hunting All the Hidden
  Photons}}, \href{https://doi.org/10.1007/JHEP07(2018)094}{\emph{JHEP}
  {\bfseries 07} (2018) 094}
  [\href{https://arxiv.org/abs/1803.05466}{{\ttfamily 1803.05466}}].

\bibitem{Han:2015yys}
T.~Han, S.~K. Kang and J.~Sayre, \emph{{Muon $g-2$ in the aligned two Higgs
  doublet model}}, \href{https://doi.org/10.1007/JHEP02(2016)097}{\emph{JHEP}
  {\bfseries 02} (2016) 097}
  [\href{https://arxiv.org/abs/1511.05162}{{\ttfamily 1511.05162}}].

\bibitem{Ilisie:2015tra}
V.~Ilisie, \emph{{New Barr-Zee contributions to $\mathbf{(g-2)_\mu}$ in
  two-Higgs-doublet models}},
  \href{https://doi.org/10.1007/JHEP04(2015)077}{\emph{JHEP} {\bfseries 04}
  (2015) 077} [\href{https://arxiv.org/abs/1502.04199}{{\ttfamily
  1502.04199}}].

\bibitem{Cherchiglia:2016eui}
A.~Cherchiglia, P.~Kneschke, D.~Stöckinger and H.~Stöckinger-Kim, \emph{{The
  muon magnetic moment in the 2HDM: complete two-loop result}},
  \href{https://doi.org/10.1007/JHEP01(2017)007}{\emph{JHEP} {\bfseries 01}
  (2017) 007} [\href{https://arxiv.org/abs/1607.06292}{{\ttfamily
  1607.06292}}].

\bibitem{Abe:2017jqo}
T.~Abe, R.~Sato and K.~Yagyu, \emph{{Muon specific two-Higgs-doublet model}},
  \href{https://doi.org/10.1007/JHEP07(2017)012}{\emph{JHEP} {\bfseries 07}
  (2017) 012} [\href{https://arxiv.org/abs/1705.01469}{{\ttfamily
  1705.01469}}].

\bibitem{Cao:2009as}
J.~Cao, P.~Wan, L.~Wu and J.~M. Yang, \emph{{Lepton-Specific Two-Higgs Doublet
  Model: Experimental Constraints and Implication on Higgs Phenomenology}},
  \href{https://doi.org/10.1103/PhysRevD.80.071701}{\emph{Phys. Rev.}
  {\bfseries D80} (2009) 071701}
  [\href{https://arxiv.org/abs/0909.5148}{{\ttfamily 0909.5148}}].

\bibitem{Broggio:2014mna}
A.~Broggio, E.~J. Chun, M.~Passera, K.~M. Patel and S.~K. Vempati,
  \emph{{Limiting two-Higgs-doublet models}},
  \href{https://doi.org/10.1007/JHEP11(2014)058}{\emph{JHEP} {\bfseries 11}
  (2014) 058} [\href{https://arxiv.org/abs/1409.3199}{{\ttfamily 1409.3199}}].

\bibitem{Wang:2014sda}
L.~Wang and X.-F. Han, \emph{{A light pseudoscalar of 2HDM confronted with muon
  g-2 and experimental constraints}},
  \href{https://doi.org/10.1007/JHEP05(2015)039}{\emph{JHEP} {\bfseries 05}
  (2015) 039} [\href{https://arxiv.org/abs/1412.4874}{{\ttfamily 1412.4874}}].

\bibitem{Abe:2015oca}
T.~Abe, R.~Sato and K.~Yagyu, \emph{{Lepton-specific two Higgs doublet model as
  a solution of muon $g-2$ anomaly}},
  \href{https://doi.org/10.1007/JHEP07(2015)064}{\emph{JHEP} {\bfseries 07}
  (2015) 064} [\href{https://arxiv.org/abs/1504.07059}{{\ttfamily
  1504.07059}}].

\bibitem{Li:2018aov}
S.-P. Li, X.-Q. Li and Y.-D. Yang, \emph{{Muon $g-2$ in a $U(1)$-symmetric
  Two-Higgs-Doublet Model}},
  \href{https://doi.org/10.1103/PhysRevD.99.035010}{\emph{Phys. Rev.}
  {\bfseries D99} (2019) 035010}
  [\href{https://arxiv.org/abs/1808.02424}{{\ttfamily 1808.02424}}].

\bibitem{Iguro:2019sly}
S.~Iguro, Y.~Omura and M.~Takeuchi, \emph{{Testing the 2HDM explanation of the
  muon g-2 anomaly at the LHC}},
  \href{https://arxiv.org/abs/1907.09845}{{\ttfamily 1907.09845}}.

\bibitem{Stockinger:2006zn}
D.~Stockinger, \emph{{The Muon Magnetic Moment and Supersymmetry}},
  \href{https://doi.org/10.1088/0954-3899/34/2/R01}{\emph{J. Phys.} {\bfseries
  G34} (2007) R45} [\href{https://arxiv.org/abs/hep-ph/0609168}{{\ttfamily
  hep-ph/0609168}}].

\bibitem{Martin:2001st}
S.~P. Martin and J.~D. Wells, \emph{{Muon Anomalous Magnetic Dipole Moment in
  Supersymmetric Theories}},
  \href{https://doi.org/10.1103/PhysRevD.64.035003}{\emph{Phys. Rev.}
  {\bfseries D64} (2001) 035003}
  [\href{https://arxiv.org/abs/hep-ph/0103067}{{\ttfamily hep-ph/0103067}}].

\bibitem{Belyaev:2016oxy}
A.~S. Belyaev, J.~E. Camargo-Molina, S.~F. King, D.~J. Miller, A.~P. Morais and
  P.~B. Schaefers, \emph{{A to Z of the Muon Anomalous Magnetic Moment in the
  MSSM with Pati-Salam at the GUT scale}},
  \href{https://doi.org/10.1007/JHEP06(2016)142}{\emph{JHEP} {\bfseries 06}
  (2016) 142} [\href{https://arxiv.org/abs/1605.02072}{{\ttfamily
  1605.02072}}].

\bibitem{Choudhury:2017fuu}
A.~Choudhury, L.~Darmé, L.~Roszkowski, E.~M. Sessolo and S.~Trojanowski,
  \emph{{Muon $g-2$ and related phenomenology in constrained vector-like
  extensions of the MSSM}},
  \href{https://doi.org/10.1007/JHEP05(2017)072}{\emph{JHEP} {\bfseries 05}
  (2017) 072} [\href{https://arxiv.org/abs/1701.08778}{{\ttfamily
  1701.08778}}].

\bibitem{Choudhury:2017acn}
A.~Choudhury, S.~Rao and L.~Roszkowski, \emph{{Impact of LHC data on muon $g-2$
  solutions in a vectorlike extension of the constrained MSSM}},
  \href{https://doi.org/10.1103/PhysRevD.96.075046}{\emph{Phys. Rev.}
  {\bfseries D96} (2017) 075046}
  [\href{https://arxiv.org/abs/1708.05675}{{\ttfamily 1708.05675}}].

\bibitem{Kotlarski:2019muo}
W.~Kotlarski, D.~Stöckinger and H.~Stöckinger-Kim, \emph{{Low-energy lepton
  physics in the MRSSM: $(g-2)_\mu$, $\mu \to e\gamma$ and $\mu\to e$
  conversion}}, \href{https://doi.org/10.1007/JHEP08(2019)082}{\emph{JHEP}
  {\bfseries 08} (2019) 082}
  [\href{https://arxiv.org/abs/1902.06650}{{\ttfamily 1902.06650}}].

\bibitem{Czarnecki:2001pv}
A.~Czarnecki and W.~J. Marciano, \emph{{The Muon anomalous magnetic moment: A
  Harbinger for 'new physics'}},
  \href{https://doi.org/10.1103/PhysRevD.64.013014}{\emph{Phys. Rev.}
  {\bfseries D64} (2001) 013014}
  [\href{https://arxiv.org/abs/hep-ph/0102122}{{\ttfamily hep-ph/0102122}}].

\bibitem{Endo:2017zrj}
M.~Endo, K.~Hamaguchi, S.~Iwamoto and K.~Yanagi, \emph{{Probing minimal SUSY
  scenarios in the light of muon $g-2$ and dark matter}},
  \href{https://doi.org/10.1007/JHEP06(2017)031}{\emph{JHEP} {\bfseries 06}
  (2017) 031} [\href{https://arxiv.org/abs/1704.05287}{{\ttfamily
  1704.05287}}].

\bibitem{Hagiwara:2017lse}
K.~Hagiwara, K.~Ma and S.~Mukhopadhyay, \emph{{Closing in on the chargino
  contribution to the muon g-2 in the MSSM: current LHC constraints}},
  \href{https://doi.org/10.1103/PhysRevD.97.055035}{\emph{Phys. Rev.}
  {\bfseries D97} (2018) 055035}
  [\href{https://arxiv.org/abs/1706.09313}{{\ttfamily 1706.09313}}].

\bibitem{Ibe:2019jbx}
M.~Ibe, M.~Suzuki, T.~T. Yanagida and N.~Yokozaki, \emph{{Muon $g-2$ in
  Split-Family SUSY in light of LHC Run II}},
  \href{https://arxiv.org/abs/1903.12433}{{\ttfamily 1903.12433}}.

\bibitem{Tran:2018kxv}
H.~M. Tran and H.~T. Nguyen, \emph{{GUT-inspired MSSM in light of muon $g-2$
  and LHC results at $\sqrt{s}=13$ TeV}},
  \href{https://doi.org/10.1103/PhysRevD.99.035040}{\emph{Phys. Rev.}
  {\bfseries D99} (2019) 035040}
  [\href{https://arxiv.org/abs/1812.11757}{{\ttfamily 1812.11757}}].

\bibitem{Endo:2020mqz}
M.~Endo, K.~Hamaguchi, S.~Iwamoto and T.~Kitahara, \emph{{Muon $g$-2 vs LHC Run
  2 in Supersymmetric Models}},
  \href{https://arxiv.org/abs/2001.11025}{{\ttfamily 2001.11025}}.

\bibitem{Cao:2016cnv}
J.~Cao, Y.~He, L.~Shang, W.~Su, P.~Wu and Y.~Zhang, \emph{{Strong constraints
  of LUX-2016 results on the natural NMSSM}},
  \href{https://doi.org/10.1007/JHEP10(2016)136}{\emph{JHEP} {\bfseries 10}
  (2016) 136} [\href{https://arxiv.org/abs/1609.00204}{{\ttfamily
  1609.00204}}].

\bibitem{Cao:2016nix}
J.~Cao, Y.~He, L.~Shang, W.~Su and Y.~Zhang, \emph{{Natural NMSSM after LHC Run
  I and the Higgsino dominated dark matter scenario}},
  \href{https://doi.org/10.1007/JHEP08(2016)037}{\emph{JHEP} {\bfseries 08}
  (2016) 037} [\href{https://arxiv.org/abs/1606.04416}{{\ttfamily
  1606.04416}}].

\bibitem{Cao:2018rix}
J.~Cao, Y.~He, L.~Shang, Y.~Zhang and P.~Zhu, \emph{{Current status of a
  natural NMSSM in light of LHC 13 TeV data and XENON-1T results}},
  \href{https://doi.org/10.1103/PhysRevD.99.075020}{\emph{Phys. Rev.}
  {\bfseries D99} (2019) 075020}
  [\href{https://arxiv.org/abs/1810.09143}{{\ttfamily 1810.09143}}].

\bibitem{Barbieri:2000gf}
R.~Barbieri and A.~Strumia, \emph{{The 'LEP paradox'}},  in \emph{{4th
  Rencontres du Vietnam: Physics at Extreme Energies (Particle Physics and
  Astrophysics) Hanoi, Vietnam, July 19-25, 2000}}, 2000,
  \href{https://arxiv.org/abs/hep-ph/0007265}{{\ttfamily hep-ph/0007265}}.

\bibitem{Giudice:2006sn}
G.~F. Giudice and R.~Rattazzi, \emph{{Living Dangerously with Low-Energy
  Supersymmetry}},
  \href{https://doi.org/10.1016/j.nuclphysb.2006.07.031}{\emph{Nucl. Phys.}
  {\bfseries B757} (2006) 19}
  [\href{https://arxiv.org/abs/hep-ph/0606105}{{\ttfamily hep-ph/0606105}}].

\bibitem{Yanagida:2017dao}
T.~T. Yanagida and N.~Yokozaki, \emph{{Muon $g-2$ in MSSM gauge mediation
  revisited}},
  \href{https://doi.org/10.1016/j.physletb.2017.07.002}{\emph{Phys. Lett.}
  {\bfseries B772} (2017) 409}
  [\href{https://arxiv.org/abs/1704.00711}{{\ttfamily 1704.00711}}].

\bibitem{Wang:2017vxj}
F.~Wang, W.~Wang and J.~M. Yang, \emph{{Solving the muon $g-2$ anomaly in
  deflected anomaly mediated SUSY breaking with messenger-matter
  interactions}}, \href{https://doi.org/10.1103/PhysRevD.96.075025}{\emph{Phys.
  Rev.} {\bfseries D96} (2017) 075025}
  [\href{https://arxiv.org/abs/1703.10894}{{\ttfamily 1703.10894}}].

\bibitem{Endo:2011xq}
M.~Endo, K.~Hamaguchi, S.~Iwamoto and N.~Yokozaki, \emph{{Higgs mass, muon
  $g-2$, and LHC prospects in gauge mediation models with vector-like
  matters}}, \href{https://doi.org/10.1103/PhysRevD.85.095012}{\emph{Phys.
  Rev.} {\bfseries D85} (2012) 095012}
  [\href{https://arxiv.org/abs/1112.5653}{{\ttfamily 1112.5653}}].

\bibitem{Abe:2002eq}
N.~Abe and M.~Endo, \emph{{Recent muon $g-2$ result in deflected anomaly
  mediated supersymmetry breaking}},
  \href{https://doi.org/10.1016/S0370-2693(03)00658-0}{\emph{Phys. Lett.}
  {\bfseries B564} (2003) 73}
  [\href{https://arxiv.org/abs/hep-ph/0212002}{{\ttfamily hep-ph/0212002}}].

\bibitem{Gogoladze:2015jua}
I.~Gogoladze, Q.~Shafi and C.~S. Ün, \emph{{Reconciling the muon $g-2$ , a 125
  GeV Higgs boson, and dark matter in gauge mediation models}},
  \href{https://doi.org/10.1103/PhysRevD.92.115014}{\emph{Phys. Rev.}
  {\bfseries D92} (2015) 115014}
  [\href{https://arxiv.org/abs/1509.07906}{{\ttfamily 1509.07906}}].

\bibitem{Ajaib:2015ika}
M.~Adeel~Ajaib, I.~Gogoladze and Q.~Shafi, \emph{{GUT-inspired supersymmetric
  model for $h\to\gamma\gamma$ and the muon $g-2$}},
  \href{https://doi.org/10.1103/PhysRevD.91.095005}{\emph{Phys. Rev.}
  {\bfseries D91} (2015) 095005}
  [\href{https://arxiv.org/abs/1501.04125}{{\ttfamily 1501.04125}}].

\bibitem{Ajaib:2014ana}
M.~A. Ajaib, I.~Gogoladze, Q.~Shafi and C.~S. Ün, \emph{{Split sfermion
  families, Yukawa unification and muon $g - 2$}},
  \href{https://doi.org/10.1007/JHEP05(2014)079}{\emph{JHEP} {\bfseries 05}
  (2014) 079} [\href{https://arxiv.org/abs/1402.4918}{{\ttfamily 1402.4918}}].

\bibitem{Babu:2014lwa}
K.~S. Babu, I.~Gogoladze, Q.~Shafi and C.~S. Ün, \emph{{Muon $g-2$, 125 GeV
  Higgs boson, and neutralino dark matter in a flavor symmetry-based MSSM}},
  \href{https://doi.org/10.1103/PhysRevD.90.116002}{\emph{Phys. Rev.}
  {\bfseries D90} (2014) 116002}
  [\href{https://arxiv.org/abs/1406.6965}{{\ttfamily 1406.6965}}].

\bibitem{Gogoladze:2014cha}
I.~Gogoladze, F.~Nasir, Q.~Shafi and C.~S. Un, \emph{{Nonuniversal Gaugino
  Masses and Muon g-2}},
  \href{https://doi.org/10.1103/PhysRevD.90.035008}{\emph{Phys. Rev.}
  {\bfseries D90} (2014) 035008}
  [\href{https://arxiv.org/abs/1403.2337}{{\ttfamily 1403.2337}}].

\bibitem{Okada:2013ija}
N.~Okada, S.~Raza and Q.~Shafi, \emph{{Particle Spectroscopy of Supersymmetric
  SU(5) in Light of 125 GeV Higgs and Muon $g-2$ Data}},
  \href{https://doi.org/10.1103/PhysRevD.90.015020}{\emph{Phys. Rev.}
  {\bfseries D90} (2014) 015020}
  [\href{https://arxiv.org/abs/1307.0461}{{\ttfamily 1307.0461}}].

\bibitem{Kowalska:2015zja}
K.~Kowalska, L.~Roszkowski, E.~M. Sessolo and A.~J. Williams,
  \emph{{GUT-inspired SUSY and the muon $g-2$ anomaly: prospects for LHC 14
  TeV}}, \href{https://doi.org/10.1007/JHEP06(2015)020}{\emph{JHEP} {\bfseries
  06} (2015) 020} [\href{https://arxiv.org/abs/1503.08219}{{\ttfamily
  1503.08219}}].

\bibitem{Mohanty:2013soa}
S.~Mohanty, S.~Rao and D.~P. Roy, \emph{{Reconciling the muon $g-2$ and dark
  matter relic density with the LHC results in nonuniversal gaugino mass
  models}}, \href{https://doi.org/10.1007/JHEP09(2013)027}{\emph{JHEP}
  {\bfseries 09} (2013) 027} [\href{https://arxiv.org/abs/1303.5830}{{\ttfamily
  1303.5830}}].

\bibitem{Akula:2013ioa}
S.~Akula and P.~Nath, \emph{{Gluino-driven radiative breaking, Higgs boson
  mass, muon $g-2$, and the Higgs diphoton decay in supergravity unification}},
  \href{https://doi.org/10.1103/PhysRevD.87.115022}{\emph{Phys. Rev.}
  {\bfseries D87} (2013) 115022}
  [\href{https://arxiv.org/abs/1304.5526}{{\ttfamily 1304.5526}}].

\bibitem{Ibe:2013oha}
M.~Ibe, T.~T. Yanagida and N.~Yokozaki, \emph{{Muon $g-2$ and 125 GeV Higgs in
  Split-Family Supersymmetry}},
  \href{https://doi.org/10.1007/JHEP08(2013)067}{\emph{JHEP} {\bfseries 08}
  (2013) 067} [\href{https://arxiv.org/abs/1303.6995}{{\ttfamily 1303.6995}}].

\bibitem{Wang:2015rli}
F.~Wang, W.~Wang and J.~M. Yang, \emph{{Reconcile muon g-2 anomaly with LHC
  data in SUGRA with generalized gravity mediation}},
  \href{https://doi.org/10.1007/JHEP06(2015)079}{\emph{JHEP} {\bfseries 06}
  (2015) 079} [\href{https://arxiv.org/abs/1504.00505}{{\ttfamily
  1504.00505}}].

\bibitem{Wang:2015nra}
F.~Wang, W.~Wang, J.~M. Yang and Y.~Zhang, \emph{{Heavy colored SUSY partners
  from deflected anomaly mediation}},
  \href{https://doi.org/10.1007/JHEP07(2015)138}{\emph{JHEP} {\bfseries 07}
  (2015) 138} [\href{https://arxiv.org/abs/1505.02785}{{\ttfamily
  1505.02785}}].

\bibitem{Bagnaschi:2017tru}
E.~Bagnaschi et~al., \emph{{Likelihood Analysis of the pMSSM11 in Light of LHC
  13-TeV Data}},
  \href{https://doi.org/10.1140/epjc/s10052-018-5697-0}{\emph{Eur. Phys. J.}
  {\bfseries C78} (2018) 256}
  [\href{https://arxiv.org/abs/1710.11091}{{\ttfamily 1710.11091}}].

\bibitem{Deppisch:2004fa}
F.~Deppisch and J.~W.~F. Valle, \emph{{Enhanced lepton flavor violation in the
  supersymmetric inverse seesaw model}},
  \href{https://doi.org/10.1103/PhysRevD.72.036001}{\emph{Phys. Rev.}
  {\bfseries D72} (2005) 036001}
  [\href{https://arxiv.org/abs/hep-ph/0406040}{{\ttfamily hep-ph/0406040}}].

\bibitem{Arina:2008bb}
C.~Arina, F.~Bazzocchi, N.~Fornengo, J.~C. Romao and J.~W.~F. Valle,
  \emph{{Minimal supergravity sneutrino dark matter and inverse seesaw neutrino
  masses}}, \href{https://doi.org/10.1103/PhysRevLett.101.161802}{\emph{Phys.
  Rev. Lett.} {\bfseries 101} (2008) 161802}
  [\href{https://arxiv.org/abs/0806.3225}{{\ttfamily 0806.3225}}].

\bibitem{Abada:2010ym}
A.~Abada, G.~Bhattacharyya, D.~Das and C.~Weiland, \emph{{A possible connection
  between neutrino mass generation and the lightness of a NMSSM pseudoscalar}},
  \href{https://doi.org/10.1016/j.physletb.2011.05.020}{\emph{Phys. Lett.}
  {\bfseries B700} (2011) 351}
  [\href{https://arxiv.org/abs/1011.5037}{{\ttfamily 1011.5037}}].

\bibitem{delAguila:2019mvp}
F.~Del~Aguila, J.~I. Illana, J.~M. Perez-Poyatos and J.~Santiago,
  \emph{{Inverse see-saw neutrino masses in the Littlest Higgs model with
  T-parity}}, \href{https://doi.org/10.1007/JHEP12(2019)154}{\emph{JHEP}
  {\bfseries 12} (2019) 154}
  [\href{https://arxiv.org/abs/1910.09569}{{\ttfamily 1910.09569}}].

\bibitem{Cao:2017cjf}
J.~Cao, X.~Guo, Y.~He, L.~Shang and Y.~Yue, \emph{{Sneutrino DM in the NMSSM
  with inverse seesaw mechanism}},
  \href{https://doi.org/10.1007/JHEP10(2017)044}{\emph{JHEP} {\bfseries 10}
  (2017) 044} [\href{https://arxiv.org/abs/1707.09626}{{\ttfamily
  1707.09626}}].

\bibitem{Cao:2019qng}
J.~Cao, L.~Meng, Y.~Yue, H.~Zhou and P.~Zhu, \emph{{Suppressing the Scattering
  of WIMP DM and Nucleon in Supersymmetric Theories}},
  \href{https://arxiv.org/abs/1910.14317}{{\ttfamily 1910.14317}}.

\bibitem{Cao:2019aam}
J.~Cao, Y.~He, L.~Meng, Y.~Pan, Y.~Yue and P.~Zhu, \emph{{Impact of leptonic
  unitary and Xenon-1T experiment on sneutrino DM physics in the NMSSM with
  Inverse-Seesaw mechanism}},
  \href{https://arxiv.org/abs/1903.01124}{{\ttfamily 1903.01124}}.

\bibitem{Baum:2017gbj}
S.~Baum, K.~Freese, N.~R. Shah and B.~Shakya, \emph{{NMSSM Higgs boson search
  strategies at the LHC and the mono-Higgs signature in particular}},
  \href{https://doi.org/10.1103/PhysRevD.95.115036}{\emph{Phys. Rev.}
  {\bfseries D95} (2017) 115036}
  [\href{https://arxiv.org/abs/1703.07800}{{\ttfamily 1703.07800}}].

\bibitem{Beskidt:2019mos}
C.~Beskidt and W.~de~Boer, \emph{{Effective scanning method of the NMSSM
  parameter space}},
  \href{https://doi.org/10.1103/PhysRevD.100.055007}{\emph{Phys. Rev.}
  {\bfseries D100} (2019) 055007}
  [\href{https://arxiv.org/abs/1905.07963}{{\ttfamily 1905.07963}}].

\bibitem{MCCALL2005205}
J.~McCall, \emph{Genetic algorithms for modelling and optimisation},
  \href{https://doi.org/https://doi.org/10.1016/j.cam.2004.07.034}{\emph{Journal
  of Computational and Applied Mathematics} {\bfseries 184} (2005) 205 }.

\bibitem{Feroz:2008xx}
F.~Feroz, M.~Hobson and M.~Bridges, \emph{{MultiNest: an efficient and robust
  Bayesian inference tool for cosmology and particle physics}},
  \href{https://doi.org/10.1111/j.1365-2966.2009.14548.x}{\emph{Mon. Not. Roy.
  Astron. Soc.} {\bfseries 398} (2009) 1601}
  [\href{https://arxiv.org/abs/0809.3437}{{\ttfamily 0809.3437}}].

\bibitem{Feroz:2013hea}
F.~Feroz, M.~Hobson, E.~Cameron and A.~Pettitt, \emph{{Importance Nested
  Sampling and the MultiNest Algorithm}},
  \href{https://arxiv.org/abs/1306.2144}{{\ttfamily 1306.2144}}.

\bibitem{Aaboud:2307399}
{\scshape ATLAS Collaboration} collaboration, \emph{{Search for electroweak
  production of supersymmetric particles in final states with two or three
  leptons at $\sqrt{s}=13$ TeV with the ATLAS detector}},
  \href{https://doi.org/10.1140/epjc/s10052-018-6423-7}{\emph{Eur. Phys. J. C}
  {\bfseries 78} (2018) 995. 36 p}.

\bibitem{Liu:2005rs}
C.~Liu, \emph{Supersymmetry for fermion masses},
  \href{https://doi.org/10.1088/0253-6102/47/6/025}{\emph{Commun.Theor.Phys.}
  {\bfseries 47} (2007) 1088}
  [\href{https://arxiv.org/abs/hep-ph/0507298}{{\ttfamily hep-ph/0507298}}].

\bibitem{Staub:2015kfa}
F.~Staub, \emph{{Exploring new models in all detail with SARAH}},
  \href{https://doi.org/10.1155/2015/840780}{\emph{Adv. High Energy Phys.}
  {\bfseries 2015} (2015) 840780}
  [\href{https://arxiv.org/abs/1503.04200}{{\ttfamily 1503.04200}}].

\bibitem{Porod:2003um}
W.~Porod, \emph{{SPheno, a program for calculating supersymmetric spectra, SUSY
  particle decays and SUSY particle production at e+ e- colliders}},
  \href{https://doi.org/10.1016/S0010-4655(03)00222-4}{\emph{Comput. Phys.
  Commun.} {\bfseries 153} (2003) 275}
  [\href{https://arxiv.org/abs/hep-ph/0301101}{{\ttfamily hep-ph/0301101}}].

\bibitem{Porod:2011nf}
W.~Porod and F.~Staub, \emph{{SPheno 3.1: Extensions including flavour,
  CP-phases and models beyond the MSSM}},
  \href{https://doi.org/10.1016/j.cpc.2012.05.021}{\emph{Comput. Phys. Commun.}
  {\bfseries 183} (2012) 2458}
  [\href{https://arxiv.org/abs/1104.1573}{{\ttfamily 1104.1573}}].

\bibitem{Belanger:2010pz}
G.~Belanger, F.~Boudjema, A.~Pukhov and A.~Semenov, \emph{micromegas: A tool
  for dark matter studies},  in \emph{micrOMEGAs: A Tool for dark matter
  studies}, vol.~033N2, pp.~111--116, 2010,
  \href{https://arxiv.org/abs/1005.4133}{{\ttfamily 1005.4133}},
  \href{https://doi.org/10.1393/ncc/i2010-10591-3}{DOI}.

\bibitem{Camargo-Molina:2013qva}
J.~E. Camargo-Molina, B.~O'Leary, W.~Porod and F.~Staub,
  \emph{{$\mathbf{Vevacious}$: A Tool For Finding The Global Minima Of One-Loop
  Effective Potentials With Many Scalars}},
  \href{https://doi.org/10.1140/epjc/s10052-013-2588-2}{\emph{Eur. Phys. J.}
  {\bfseries C73} (2013) 2588}
  [\href{https://arxiv.org/abs/1307.1477}{{\ttfamily 1307.1477}}].

\bibitem{Camargo-Molina:2014pwa}
J.~E. Camargo-Molina, B.~Garbrecht, B.~O'Leary, W.~Porod and F.~Staub,
  \emph{{Constraining the Natural MSSM through tunneling to color-breaking
  vacua at zero and non-zero temperature}},
  \href{https://doi.org/10.1016/j.physletb.2014.08.036}{\emph{Phys. Lett.}
  {\bfseries B737} (2014) 156}
  [\href{https://arxiv.org/abs/1405.7376}{{\ttfamily 1405.7376}}].

\bibitem{Wainwright:2011kj}
C.~L. Wainwright, \emph{{CosmoTransitions: Computing Cosmological Phase
  Transition Temperatures and Bubble Profiles with Multiple Fields}},
  \href{https://doi.org/10.1016/j.cpc.2012.04.004}{\emph{Comput. Phys. Commun.}
  {\bfseries 183} (2012) 2006}
  [\href{https://arxiv.org/abs/1109.4189}{{\ttfamily 1109.4189}}].

\bibitem{Sirunyan:2272260}
{\scshape CMS Collaboration} collaboration, \emph{{Measurements of properties
  of the Higgs boson decaying into the four-lepton final state in pp collisions
  at sqrt(s) = 13 TeV. Measurements of properties of the Higgs boson decaying
  into the four-lepton final state in pp collisions at sqrt(s) = 13 TeV}},
  \href{https://doi.org/10.1007/JHEP11(2017)047}{\emph{JHEP} {\bfseries 11}
  (2017) 047. 55 p}.

\bibitem{ATLAS-CONF-2017-046}
{\scshape ATLAS Collaboration} collaboration, \emph{{Measurement of the Higgs
  boson mass in the $H\rightarrow ZZ^*\rightarrow 4\ell$ and
  $H\rightarrow\gamma\gamma$ channels with $\sqrt{s}$=13TeV $pp$ collisions
  using the ATLAS detector}},  Tech. Rep. ATLAS-CONF-2017-046, CERN, Geneva,
  Jul, 2017.

\bibitem{Bechtle:2008jh}
P.~Bechtle, O.~Brein, S.~Heinemeyer, G.~Weiglein and K.~E. Williams,
  \emph{{HiggsBounds: Confronting Arbitrary Higgs Sectors with Exclusion Bounds
  from LEP and the Tevatron}},
  \href{https://doi.org/10.1016/j.cpc.2009.09.003}{\emph{Comput. Phys. Commun.}
  {\bfseries 181} (2010) 138}
  [\href{https://arxiv.org/abs/0811.4169}{{\ttfamily 0811.4169}}].

\bibitem{Bechtle:2011sb}
P.~Bechtle, O.~Brein, S.~Heinemeyer, G.~Weiglein and K.~E. Williams,
  \emph{{HiggsBounds 2.0.0: Confronting Neutral and Charged Higgs Sector
  Predictions with Exclusion Bounds from LEP and the Tevatron}},
  \href{https://doi.org/10.1016/j.cpc.2011.07.015}{\emph{Comput. Phys. Commun.}
  {\bfseries 182} (2011) 2605}
  [\href{https://arxiv.org/abs/1102.1898}{{\ttfamily 1102.1898}}].

\bibitem{Bechtle:2013xfa}
P.~Bechtle, S.~Heinemeyer, O.~Stål, T.~Stefaniak and G.~Weiglein,
  \emph{{$HiggsSignals$: Confronting arbitrary Higgs sectors with measurements
  at the Tevatron and the LHC}},
  \href{https://doi.org/10.1140/epjc/s10052-013-2711-4}{\emph{Eur. Phys. J.}
  {\bfseries C74} (2014) 2711}
  [\href{https://arxiv.org/abs/1305.1933}{{\ttfamily 1305.1933}}].

\bibitem{Bechtle:2014ewa}
P.~Bechtle, S.~Heinemeyer, O.~Stål, T.~Stefaniak and G.~Weiglein,
  \emph{{Probing the Standard Model with Higgs signal rates from the Tevatron,
  the LHC and a future ILC}},
  \href{https://doi.org/10.1007/JHEP11(2014)039}{\emph{JHEP} {\bfseries 11}
  (2014) 039} [\href{https://arxiv.org/abs/1403.1582}{{\ttfamily 1403.1582}}].

\bibitem{Aghanim:2018eyx}
{\scshape Planck} collaboration, N.~Aghanim et~al., \emph{{Planck 2018 results.
  VI. Cosmological parameters}},
  \href{https://arxiv.org/abs/1807.06209}{{\ttfamily 1807.06209}}.

\bibitem{Aprile:2018dbl}
{\scshape XENON} collaboration, E.~Aprile et~al., \emph{{Dark Matter Search
  Results from a One Ton-Year Exposure of XENON1T}},
  \href{https://doi.org/10.1103/PhysRevLett.121.111302}{\emph{Phys. Rev. Lett.}
  {\bfseries 121} (2018) 111302}
  [\href{https://arxiv.org/abs/1805.12562}{{\ttfamily 1805.12562}}].

\bibitem{Fuks:2013vua}
B.~Fuks, M.~Klasen, D.~R. Lamprea and M.~Rothering, \emph{{Precision
  predictions for electroweak superpartner production at hadron colliders with
  Resummino}}, \href{https://doi.org/10.1140/epjc/s10052-013-2480-0}{\emph{Eur.
  Phys. J.} {\bfseries C73} (2013) 2480}
  [\href{https://arxiv.org/abs/1304.0790}{{\ttfamily 1304.0790}}].

\bibitem{LHC-SUSY-Xsect}
{\scshape LHC SUSY Cross Section Working Group} collaboration, C.~Borschensky,
  Z.~Gecse, M.~Kraemer, R.~van~der Leeuw, A.~Kulesza, M.~Mangano et~al.,
  ``Cross sections for various subprocesses-13 tev.''
  \url{https://twiki.cern.ch/twiki/bin/view/LHCPhysics/SUSYCrossSections#Cross_sections_for_various_s_AN2}.

\bibitem{ATLAS-CONF-2019-008}
{\scshape ATLAS Collaboration} collaboration, \emph{{Search for electroweak
  production of charginos and sleptons decaying in final states with two
  leptons and missing transverse momentum in $\sqrt{s}=13$ TeV $pp$ collisions
  using the ATLAS detector}},  Tech. Rep. ATLAS-CONF-2019-008, CERN, Geneva,
  Mar, 2019.

\bibitem{Hanneke:2008tm}
D.~Hanneke, S.~Fogwell and G.~Gabrielse, \emph{{New Measurement of the Electron
  Magnetic Moment and the Fine Structure Constant}},
  \href{https://doi.org/10.1103/PhysRevLett.100.120801}{\emph{Phys. Rev. Lett.}
  {\bfseries 100} (2008) 120801}
  [\href{https://arxiv.org/abs/0801.1134}{{\ttfamily 0801.1134}}].

\bibitem{Hanneke:2010au}
D.~Hanneke, S.~F. Hoogerheide and G.~Gabrielse, \emph{{Cavity Control of a
  Single-Electron Quantum Cyclotron: Measuring the Electron Magnetic Moment}},
  \href{https://doi.org/10.1103/PhysRevA.83.052122}{\emph{Phys. Rev.}
  {\bfseries A83} (2011) 052122}
  [\href{https://arxiv.org/abs/1009.4831}{{\ttfamily 1009.4831}}].

\bibitem{Badziak:2019gaf}
M.~Badziak and K.~Sakurai, \emph{{Explanation of electron and muon g $-$ 2
  anomalies in the MSSM}},
  \href{https://doi.org/10.1007/JHEP10(2019)024}{\emph{JHEP} {\bfseries 10}
  (2019) 024} [\href{https://arxiv.org/abs/1908.03607}{{\ttfamily
  1908.03607}}].

\end{thebibliography}
\end{document}